\documentclass[aps,prd,reprint,twocolumn,10pt,eqsecnum,showpacs,nofootinbib,amsfonts,amssymb,amsmath,longbibliography,superscriptaddress,floatfix]{revtex4-1}
\usepackage[colorlinks=true,allcolors=blue, pdfusetitle]{hyperref}
\usepackage[all]{hypcap}
\usepackage{bm}
\usepackage{mathrsfs}
\usepackage[dvipsnames]{xcolor}
\usepackage{mathtools}
\usepackage{graphicx}
\usepackage{orcidlink}
\usepackage{tensor}

\graphicspath{{../plots/}}

\usepackage[T1]{fontenc}
\usepackage{lmodern}
\usepackage[utf8]{inputenc}
\usepackage{microtype}

\DeclareMathAlphabet{\mathbfsf}{\encodingdefault}{\sfdefault}{bx}{sl}

\usepackage[normalem]{ulem}

\newcommand{\Caltech}{\affiliation{Theoretical Astrophysics, Walter Burke Institute for Theoretical Physics, California Institute of Technology, Pasadena, California 91125, USA}}
\newcommand{\CornellPhysics}{\affiliation{Department of Physics, Cornell University, Ithaca, NY, 14853, USA}}
\newcommand{\Cornell}{\affiliation{Cornell Center for Astrophysics and Planetary Science, Cornell University, Ithaca, New York 14853, USA}}
\newcommand\UMiss{\affiliation{Department of Physics and Astronomy, University of Mississippi, University, MS 38677, USA}}
\newcommand{\AEI}{\affiliation{Max Planck Institute for Gravitational Physics (Albert Einstein Institute), Am M\"uhlenberg 1, D-14476 Potsdam, Germany}}

\newcommand{\figCyl}{%
	\begin{figure*}[t]
		\begin{center}
			\includegraphics[width=0.65\columnwidth]{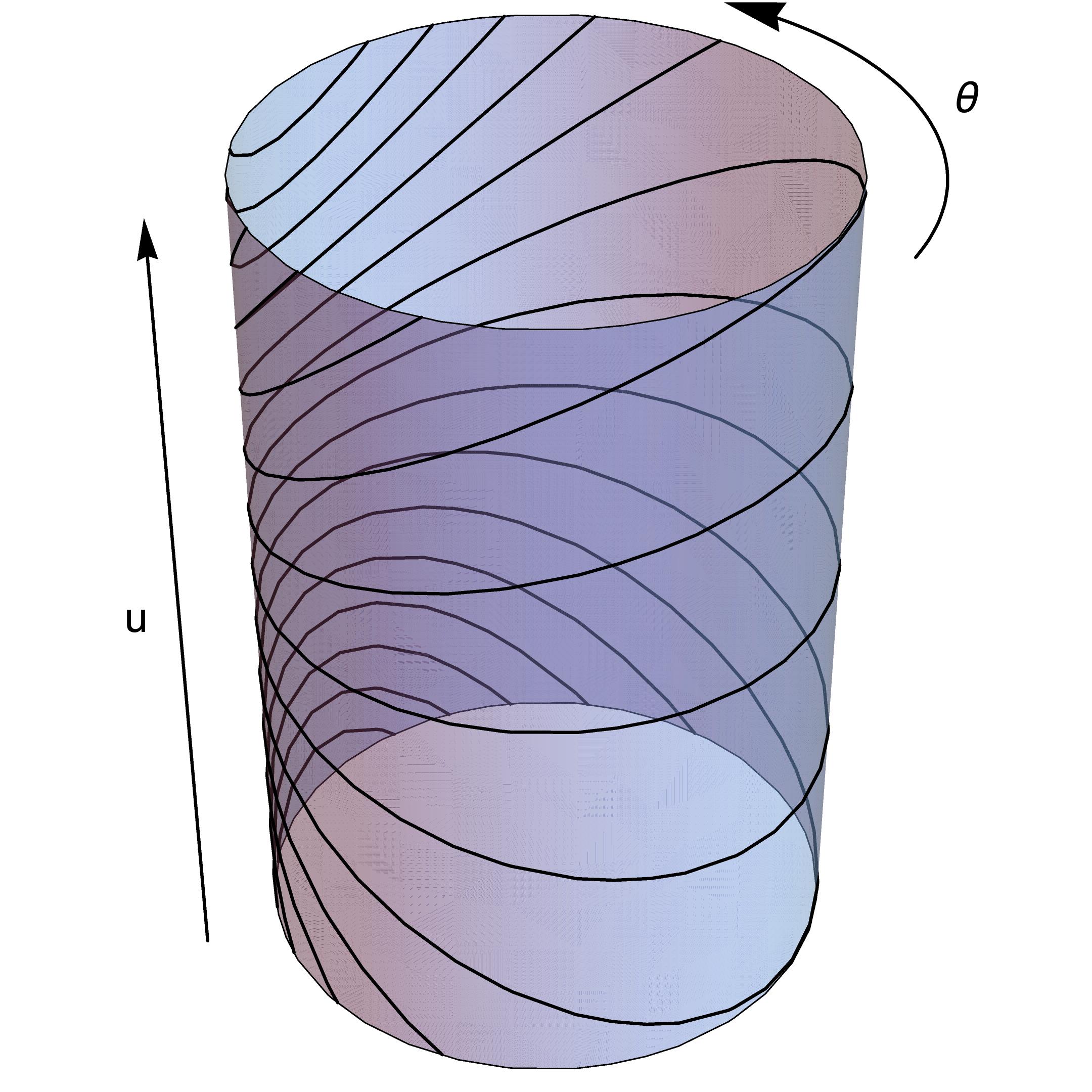}
			\includegraphics[width=0.65\columnwidth]{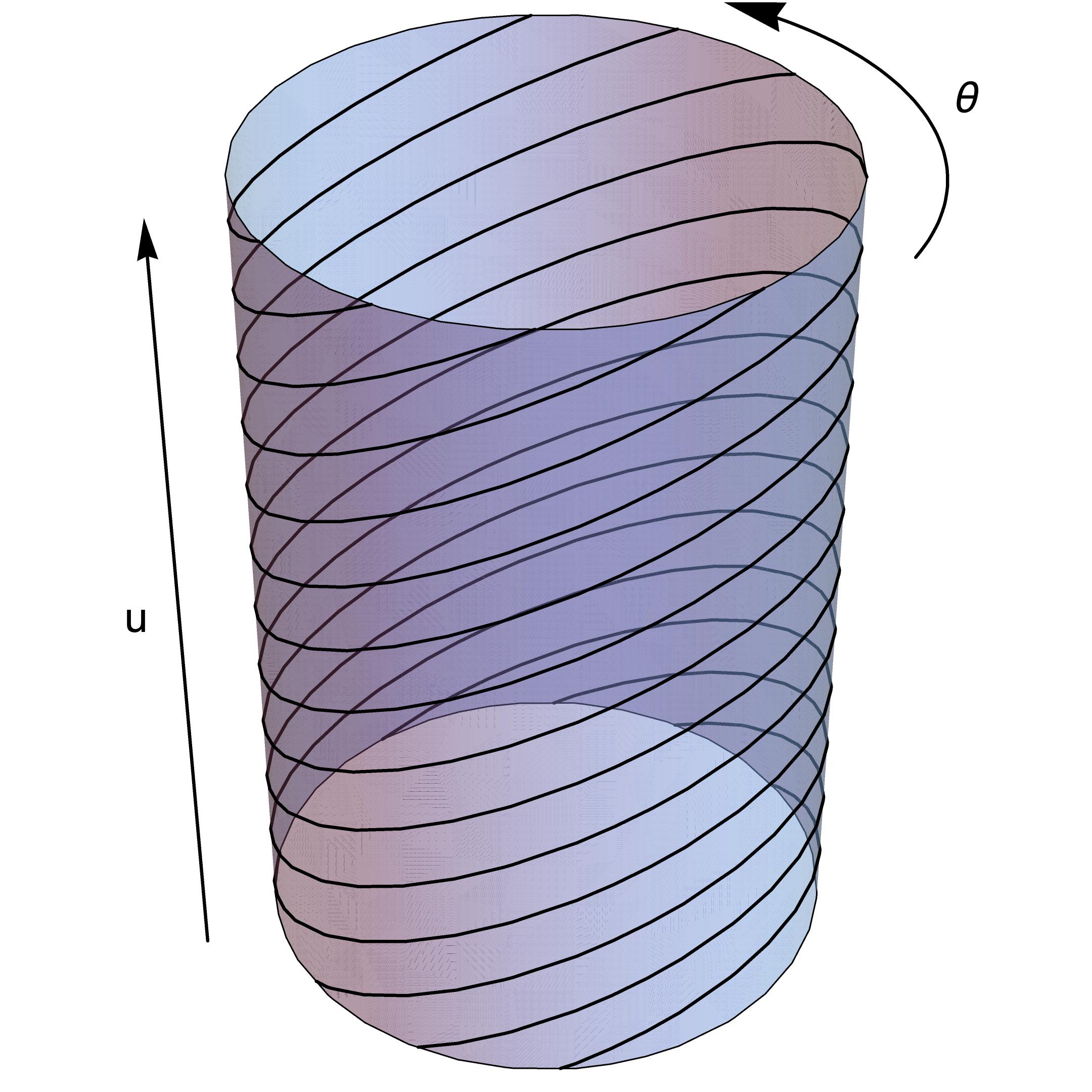}
			\includegraphics[width=0.65\columnwidth]{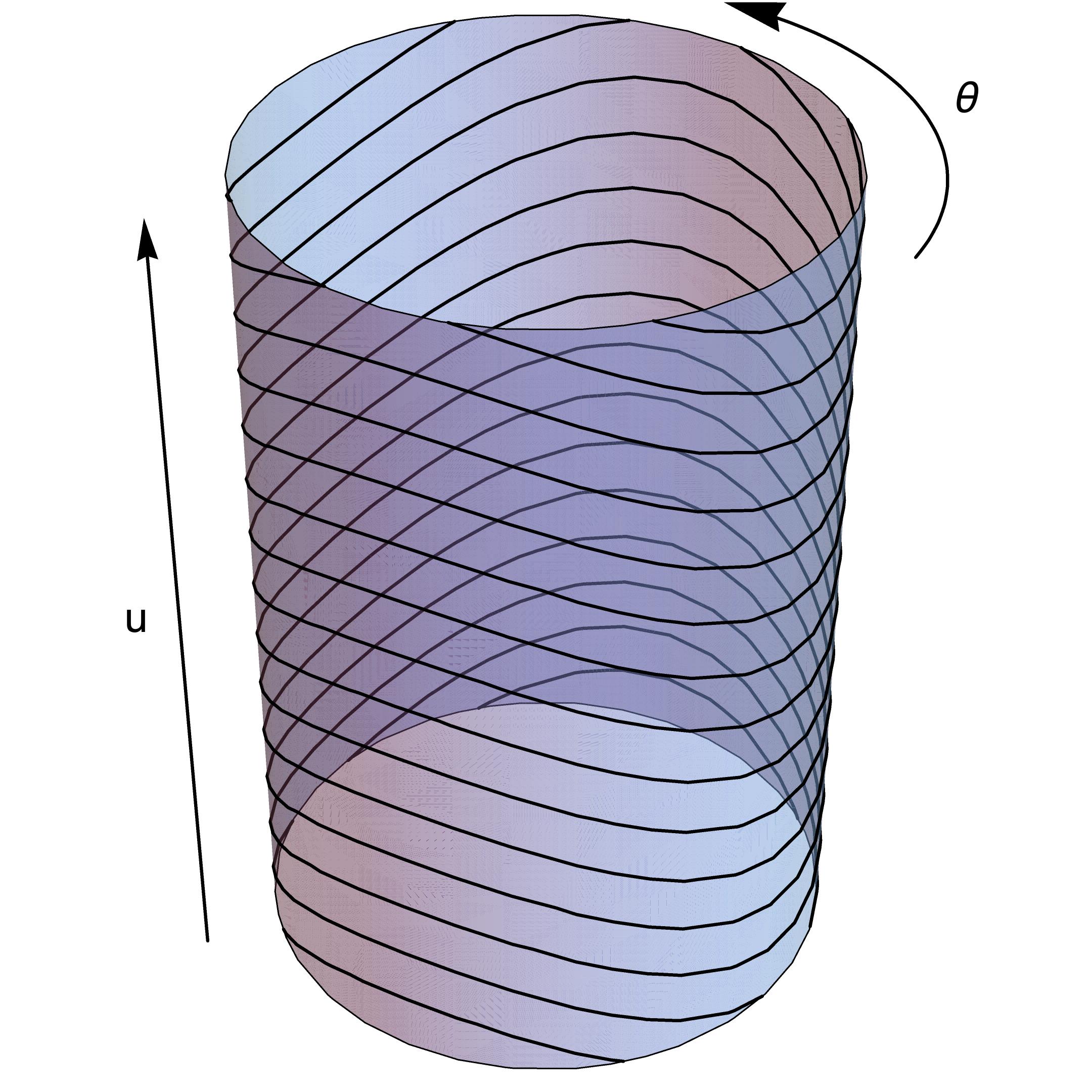}
		\end{center}
		\caption{%
			``Cylinder'' diagram of $\mathcal{I}^{+}$ to provide intuition about
			(i) boosts, (ii) space translations, and (iii) proper
      supertranslations. The retarded time coordinate $u$ runs
      vertically in each plot, while the polar coordinate $\theta$
      runs azimuthally. Black circles correspond to surfaces of constant $u$ in the untransformed frame. A Lorentz boost dilates $u$ by some factor at each $\theta$ point, while a space translation and a proper supertranslation instead shift $u$ by some function of $\theta$, which can be written as some combination of $\ell=1$ spherical harmonics for space translations, and some combination of $\ell\geq2$ spherical harmonics for proper supertranslations. The boost is in the $\hat{z}$ direction; the space translation is proportional to the spherical harmonic $Y_{(1,0)}(\theta,\phi)$, i.e., the $\hat{z}$ direction; and the proper supertranslation is proportional the spherical harmonic $Y_{(2,0)}(\theta,\phi)$.
		}
		\label{fig:Cyl}
	\end{figure*}
}

\begin{document}

\author{Keefe Mitman
  \orcidlink{0000-0003-0276-3856}}
\email{kmitman@caltech.edu} \Caltech
\author{Michael Boyle \orcidlink{0000-0002-5075-5116}} \Cornell
\author{Leo C. Stein \orcidlink{0000-0001-7559-9597}} \UMiss
\author{\\Nils Deppe \orcidlink{0000-0003-4557-4115}} \Cornell\CornellPhysics
\author{Lawrence E.~Kidder \orcidlink{0000-0001-5392-7342}} \Cornell
\author{Jordan Moxon \orcidlink{0000-0001-9009-6955}} \Caltech
\author{Harald P. Pfeiffer \orcidlink{0000-0001-9288-519X}} \AEI
\author{Mark A. Scheel \orcidlink{0000-0001-6656-9134}} \Caltech
\author{Saul A. Teukolsky \orcidlink{0000-0001-9765-4526}} \Caltech\Cornell
\author{William Throwe \orcidlink{0000-0001-5059-4378}} \Cornell
\author{Nils L. Vu \orcidlink{0000-0002-5767-3949}} \Caltech

\hypersetup{pdfauthor={Mitman et al.}}

\title{A Review of Gravitational Memory and BMS Frame Fixing in Numerical Relativity}

\begin{abstract}
  Gravitational memory effects and the BMS freedoms exhibited at future
  null infinity have recently been resolved and utilized in numerical
  relativity simulations. With this, gravitational wave models and our understanding of the fundamental nature of general relativity have been vastly improved. In this paper, we review the history and intuition
  behind memory effects and BMS symmetries, how they manifest in gravitational waves, and how controlling the infinite number of BMS freedoms of numerical relativity
  simulations can crucially improve the waveform models that are used by
  gravitational wave detectors. We reiterate the fact that, with memory effects and BMS symmetries, not only can these next-generation numerical waveforms be used to observe never-before-seen physics, but they can also be used to test GR and learn new astrophysical information about our universe.
\end{abstract}

\maketitle

\section{Introduction}
\label{sec:introduction}

One of the most pressing challenges for physics in the near future is
performing stringent and robust tests of Einstein's theory of general
relativity (GR). These tests are of the utmost importance because they
will inform us about the nature of gravity within our universe and
will reveal when our long-standing theory of GR fails to explain real
world phenomena. At present, the most prospective tests of GR that we
can perform are those which involve analyses of the gravitational
waves (GWs) that are created by binary black hole mergers
(BBHs).\footnote{%
  While the imaging of black holes, such as those performed by the EHT
  Collaboration, can also test Einstein's theory of GR, this typically probes much lower curvatures than are accessible by the LIGO-Virgo-KAGRA Collaboration and thus probe alternative, but also complementary, regimes of GR.} %
This is because the GWs that are produced by BBHs are largely
influenced by the strong-gravity regimes sourced by two coalescing
black holes and should thus capture whatever deviations from GR there
may be. However, to verify whether certain features in observed GWs
are evidence for unknown physics, we first need to have a sound
understanding of the GWs that GR predicts.

At present, the best solutions to Einstein's equations, i.e., the GW
templates used to perform tests of GR, are those produced by numerical
relativity (NR) simulations. Calculating the GWs
sourced during the coalescence of two black holes is
impossible to work out with pen and paper due to the overall complexity of the
partial differential equations that need to be solved. 

Furthermore, even if one uses perturbation theory to try to predict the GWs, this fails to produce reliable results during the merger phase of the
binary, which is typically the loudest and most detectable part of the
GW signal. Consequently, GW models that are built by using perturbation theory
or GW phenomenology, like effective one-body (EOB) or Phenomenological (Phenom)
GW models~\cite{Buonanno:1998gg,Husa:2015iqa,Khan:2015jqa,Pratten:2020ceb,Estelles:2021gvs,Taracchini:2012ig,Lackey:2013axa,Pan:2013rra,Bohe:2016gbl,Ossokine:2020kjp,Ramos-Buades:2023ehm},
always need to be calibrated against NR waveforms. Thus, NR simulations, which can achieve arbitrary
accuracy with the right computational tools, are at the heart of producing accurate and robust solutions for GW-emitting phenomena in GR.

Nonetheless, NR can still fail to accurately simulate GR if the code
infrastructure is not formulated correctly or if the necessary
numerical resolution is not achieved. One such example of this
inaccuracy was the inability of NR simulations to resolve a collection
of observables in GR colloquially referred to as memory
effects~\cite{Zeldovich:1974gvh,Braginsky:1985vlg,Favata:2010zu}. These effects are not-yet observed, nonlinear predictions of
GR that physically correspond to the net displacement that two
freely-falling observers will experience due to the passage of
transient GWs. However, apart from being a curious prediction of
GR, what makes memory effects particularly tantalizing is that they
are intimately tied---through conservation laws---to the symmetry
group of future null infinity---part of the asymptotic boundary of
spacetime. This symmetry group is not the usual
Poincaré group of special relativity, but is a larger group called the BMS group, named after Bondi, van der Burg, Metzner, and Sachs~\cite{Bondi:1962px,Sachs:1962wk}. Thus, there is hope that
with the detection of memory we can not only conduct more stringent
tests of GR, but we may even obtain a better understanding of the
asymptotic structure of our encompassing universe, which is of immense
interest to theorists trying to formulate a theory of quantum gravity
through topics such as celestial holography~\cite{Strominger:2017zoo,
  Pasterski:2019ceq, Raclariu:2021zjz, Pasterski:2021rjz,
  Pasterski:2021raf}.

In this review, we highlight recent advancements made in the NR
community to resolve memory effects as well as some work
showing how fixing the BMS freedoms at future null infinity
drastically improves both the accuracy and robustness of GW models and
analyses. Specifically, in Sec.~\ref{sec:BMSoverview} we begin
by providing some motivation for and intuition behind the BMS group
and memory effects.
Next, in
Sec.~\ref{sec:literaturereview} we provide a review of the literature
on memory, the BMS group, and BMS frame fixing. Then, in
Sec.~\ref{sec:mathoverview}, we provide a more formal explanation of
the origins of the BMS group and how memory effects can be understood
as stemming from certain conservation laws related to the
symmetries of null infinity. In
Sec.~\ref{sec:codeframeworks}, we transition to a review of
the code frameworks used to compute gravitational waves at null
infinity and we highlight the advancements in the NR community
that has made the resolution of memory effects possible. Then, in
Sec.~\ref{sec:memoryresults}, we demonstrate the formalism from
Sec.~\ref{sec:mathoverview}
using binary black hole merger simulations. In particular, we present how the
BMS conservation laws can be used to efficiently analyze gravitational
waves and understand memory effects. Furthermore, in this section we
also provide a review of the detectability of memory and the forecast
for its detection in the coming decade. In
Sec.~\ref{sec:BMSresults} we highlight the often-overlooked
importance of fixing the BMS freedom of NR waveforms to ensure that modeling of NR waveforms is performed accurately and robustly. We show
this by introducing the \emph{superrest
frame}, and
demonstrating its utility by comparing NR waveforms to post-Newtonian waveforms and by fitting NR waveforms with predictions made by black
hole perturbation theory. Finally, in Sec.~\ref{sec:discussion} we
summarize the main points of this review and provide some outlook
regarding the future of numerical relativity, memory effects, and testing the nature of gravity with gravitational waves.

\section{Pedagogical approach to\\BMS and memory}
\label{sec:BMSoverview}

Despite its importance and interesting characteristics, many relativists are not familiar with the BMS group or its effect on asymptotic data. In this section, we provide a pedagogical introduction to the BMS group, with the intention of making the rest of the paper more accessible. We begin by motivating the need for a coordinate system that is adapted to inertial observers, e.g., GW detectors, and discuss how such a coordinate system is provided by ``Bondi gauge''---in which the metric asymptotes to the usual Minkowski metric at large radius. This is crucial because GW waveforms are always studied in a certain coordinate system, so to provide meaningful waveforms we need a meaningful coordinate system that matches that of our inertial detectors. As we will see, it turns out that once such a coordinate system is constructed by mapping the metric to Bondi gauge, there is a residual ambiguity in the coordinates, i.e., a symmetry, which is described by the BMS group. The BMS group, however, is simply the usual Lorentz group, augmented by a generalized class of spacetime translations called \emph{supertranslations}~\cite{Bondi:1962px,Sachs:1962wk}. Consequently, it is fairly straightforward to understand the BMS group once the origin of and intuition behind supertranslations is understood. Thus, once we motivate the need for a meaningful coordinate system, we will then provide some intuition for supertranslations through a few informative examples involving null rays propagating in Minkowski space. Following this, we then conclude this pedagogical overview by showing how the BMS group changes the asymptotic data that can be measured by an inertial observer.  This action forms the basis for fixing the BMS frame in numerical relativity, as outlined in Sec.~\ref{sec:mathoverview}, and also provides a unique way to study memory effects, which we utilize in Sec.~\ref{sec:memoryoverview} and in the rest of the paper.

\subsection{Motivation}
\label{sec:BMSmotivation}

Choosing coordinates in GR is one of the more delicate and, at times, confusing components of Einstein's theory. In fact, for decades after GR's development, researchers---including Einstein---wavered over the issue of whether or not gravitational waves were really physically observable or simply gauge artifacts~\cite{Cervantes-Cota:2016zjc}. Ultimately, the reason for their misgivings was that GWs are often studied in terms of components of the metric or Riemann tensors, with respect to \emph{some basis determined by the coordinates},\footnote{See, e.g., Refs.~\cite{Nerozzi:2004wv, Nerozzi:2016kky} for interesting efforts to find tetrads specified by the geometry rather than arbitrary coordinates.} and \emph{expressed as functions of those coordinates}. Fortunately, this confusion regarding the observability of GWs was resolved by Pirani in 1956, who clarified their existence using tetrad methods and worldlines of particles~\cite{Pirani:1956tn}. Pirani's approach, however, was really only useful for formulating theoretical perspectives and could not be used to make statements about particular systems, like black hole mergers. For this, a more suitable framework was developed in a series of works by Bondi, van der Burg, Metzner, and Sachs~\cite{Bondi1960,Sachs1961,Bondi:1962px,Sachs:1962wk,Sachs1962PR}. Their approach to studying GWs, which we describe in Sec.~\ref{sec:Bondigauge}, involves constructing an explicit coordinate system and assuming a particular, but well-motivated asymptotic behavior of the spacetime metric in those coordinates.

As an alternative perspective regarding the subtleties of coordinates in GR, consider a numerical simulation. At the simplest level, numerical relativists must produce waveforms as tables of timestamps and corresponding GW strain values measured at various angular locations encompassing the source.\footnote{In reality, numerical relativists provide the angular dependence of the strain by representing the strain with respect to some set of angular basis functions, like spin-weighted spherical harmonics. However, to make the connection to coordinates more apparent, in this discussion we ignore this detail and instead consider the naive representation in terms of points on the two-sphere.} However, the meaning of those time coordinates, the angular locations at which the strain is measured, and the basis with respect to which the strain is evaluated rely on the essentially arbitrary coordinates used in the numerical simulation---coordinates imprinted by the vagaries of initial data and complicated gauges. Despite confusing declarations that may be found in the NR literature, no GW extraction method can produce ``invariant'' results in the sense of being independent of the choice of coordinates~\cite{Lehner:2007ip}. Even at linear order in the size of the gauge perturbation, every waveform description is coordinate-dependent. But this issue is not unique to NR simulations. Other gravitational wave modelers, such as those working in post-Newtonian (PN) or even post-Minkowskian (PM) theory, face similar ambiguities. Ideally, we would resolve these coordinate issues in a consistent way, so that waveforms from other simulations, or other models, can be compared to each other.

Due to the diffeomorphism invariance of GR, there is a rich set of coordinate systems that, in principle, could be used to study GWs. In practice, however, working with so many possible coordinate systems is not feasible. Instead, we need some well-motivated way to limit the possible coordinate systems that we can use when studying GWs. To do this, one property that we might impose is that the coordinate systems we consider be adapted to trajectories of inertial observers. That is, curves that have constant spatial coordinates could be timelike geodesics, and the time coordinate for those curves could be the proper time measured on those geodesics. While this would certainly be possible, one issue that arises is that the coordinates we consider would then depend on their initial conditions, and would surely encounter coordinate singularities. But, it turns out that if we instead consider the region of spacetime infinitely far away from the source, then it is sometimes possible\footnote{This possibility rests on some fairly stringent requirements about the spacetime, including the existence of the infinite radius limit, and the fall-off behavior of the metric in that limit. In particular, these requirements rule out direct application to, for example, FLRW spacetimes. Nonetheless, recent work has sought to extend similar analyses to FLRW spacetimes~\cite{BongaPrabhu2020, RojoSchroder2023, JokelaKajantieSarkkinen2022, RojoHeckelbacher2021, FernandezAlvarezSenovilla2022, RojoHeckelbacherOliveri2023}.} to find a set of coordinates that is \emph{asymptotically} inertial. This realization is exactly what Bondi, van der Burg, Metzner, and Sachs came to in the 1960s~\cite{Bondi1960,Sachs1961,Bondi:1962px,Sachs:1962wk,Sachs1962PR}. The core idea is that one should instead model a GW source as an isolated system, with the spacetime approaching Minkowski space far from the source, so that one can then match the coordinates to the more familiar inertial trajectories of Minkowski space.

\subsection{Bondi gauge and inertial observers}
\label{sec:Bondigauge}

The Bondi-Sachs formalism\footnote{For reasons that are not immediately apparent from the literature, various aspects of this formalism are credited to various subsets of the authors of the papers in which they first appeared: Bondi, van der Burg, Metzner, and Sachs~\cite{Bondi1960, Sachs1961, Bondi:1962px, Sachs:1962wk, Sachs1962PR}.  In particular, Bondi is credited for the gauge or frame; Bondi and Sachs for the coordinates, metric, and formalism generally; and Bondi, Metzner, and Sachs for the (BMS) group. For some reason, van der Burg seems to be left out of the conversation. In this work, we follow this convention without claiming to understand why.} begins with a collection of coordinates called Bondi-Sachs coordinates that are suited to the problem of outgoing radiation. Essentially, in this formalism we have the usual spherical coordinates $(\theta,\phi)$ as well as a null, retarded-time coordinate $u$, such that the $u$ direction is orthogonal to the $\theta$ and $\phi$ directions, and an areal coordinate $r$ relative to the $(\theta,\phi)$ coordinates. Anywhere that such a set of coordinates exists, the metric can be written in Bondi-Sachs form as
\begin{align}
\label{eq:BSmetric}
ds^{2}&=-Ue^{2\beta}du^{2}-2e^{2\beta}dudr\nonumber\\
&\phantom{=.}+r^{2}\gamma_{AB}\left(dx^{A}-\mathcal{U}^{A}du\right)\left(dx^{B}-\mathcal{U}^{B}du\right),
\end{align}
where capital Latin indices range over $(\theta,\phi)$, and we have introduced the arbitrary functions $U$, $\beta$, $\mathcal{U}^{A}$, and $\gamma_{AB}$, each of which is a function of the coordinates $(u,r,\theta,\phi)$.

With this set of intuitive coordinates, we then restrict the possible metrics that we allow by imposing certain boundary conditions, i.e., some asymptotic behavior in the limit of large radius. In particular, to ensure that the metric in this Bondi-Sachs coordinate system approaches the standard Minkowski metric in the large radius limit, we require that our metric functions obey
\begin{subequations}
\label{eq:Bondi-gauge}
\begin{align}
U&\rightarrow1,\\
\beta&\rightarrow0,\\
\mathcal{U}^{A}&\rightarrow0,\\
\label{eq:Bondi-gauge-angular}
\gamma_{AB}&\rightarrow\begin{pmatrix}1&0\\0&\sin^{2}\theta\end{pmatrix}.
\end{align}
\end{subequations}
This asymptotic restriction is exactly what is meant by being in ``Bondi gauge'' or some ``Bondi frame''. It provided an early notion of what is called ``asymptotic flatness''.

After this work of Bondi, various authors introduced important generalizations of this falloff condition~\cite{NewmanUnti1962, Penrose1963, Penrose1965, HawkingEllis1976, Geroch1977, GerochHorowitz1978}, most of which will be beyond the scope of this paper. But one that will be conceptually useful for this review is Penrose's notion of conformal compactification~\cite{Penrose1963, Penrose1965}. This compactification, which introduces extra points to construct a boundary of spacetime in the $r\rightarrow\infty$ limit, is essential for formulating ``future null infinity'' $\mathcal{I}^{+}$: the final destination of outgoing radiation.\footnote{Note that an identical description also exists at \emph{past} null infinity. But, because we will focus on future null infinity in this review, we ignore this subtlety for the remainder of this work.} This boundary is obtained by taking this $r\rightarrow\infty$ limit for fixed $u$, which yields a region of spacetime parameterized by $(u,\theta,\phi)$. When studying GWs and other asymptotic data, we will be interested in the value of this data \emph{on} this boundary. This is because, as one approaches $\mathcal{I}^{+}$, curves of constant $(r,\theta,\phi)$ are \emph{nearly} geodesics, with $u$ \emph{nearly} parametrizing the proper time, and errors in this geodesic approximation falling off as $1/r$. Consequently, GWs measured by a distant inertial observer can be approximated (at least over finite time spans) by GWs studied at $\mathcal{I}^{+}$. In fact, the Bondi frame is sometimes even referred to as the ``asymptotic inertial frame''~\cite{MadlerWinicour2016}.

Now, apart from constructing a coordinate system and a region of spacetime that can be used to study GWs that agree (up to $1/r$ corrections) with what a distant, inertial observer would see, the other import consequence of constructing the Bondi frame is that, by doing so, we have drastically reduced the coordinate ambiguity that would otherwise plague our GW waveforms. In particular, for the numerical simulation example, we no longer have to express the waveform in terms of the simulation's arbitrary coordinates, which are ambiguous up to the entire diffeomorphism group. Nor do we have to integrate a family of timelike geodesics and evaluate the waveform along those curves, still ambiguous up to the choice of initial conditions for each geodesic. Instead, we can simply express the waveform as a function of $(u,\theta,\phi)$ on $\mathcal{I}^{+}$.

But one obvious question persists: how much ambiguity still remains in this Bondi frame description? Intuitively, we might expect that the answer would simply be the transformations that leave the usual Minkowski metric unchanged, i.e., the Poincaré group. In fact, this is indeed what Bondi, van der Burg, Metzner, and Sachs thought they would find when studying the symmetry group of future null infinity. However, this intuition turns out to be only nearly correct. Specifically, the symmetry group of future null infinity is the Poincaré group, but with the usual four spacetime translations replaced by a larger set of spacetime transformations called \emph{supertranslations}. This is the BMS group.

\subsection{The BMS group}
\label{sec:BMSgroup}

The full BMS group is simply the set of transformations of the asymptotic coordinates---that is, $(u,\theta,\phi)$---that preserve the asymptotic form of the metric described in Eqs.~\eqref{eq:BSmetric} and~\eqref{eq:Bondi-gauge}.\footnote{With the condition that the angular metric $\gamma_{AB}$ must asymptote to the usual unit sphere metric, Eq.~\eqref{eq:Bondi-gauge-angular}, the asymptotic gauge conditions are preserved by the standard BMS group. However, by relaxing the condition on $\gamma_{AB}$ so that it must asymptote to anything \emph{conformally related to} the usual unit sphere metric (where the conformal factor can depend on both the angular and retarded-time coordinates), we instead obtain the ``extended''~\cite{BarnichTroessaert2010} BMS group. By relaxing the condition so that the determinant of $\gamma_{AB}$ must asymptote merely to have a specified determinant (which may be a function of $u$, $\theta$, and $\phi$), we then obtain the ``generalized''~\cite{CampigliaLaddha2014, CampigliaLaddha2015, CompereFiorucciRuzziconi2018, FlanaganPrabhuShehzad2020} BMS group. While these BMS variants are certainly interesting in some contexts, like celestial holography (see, e.g., Ref.~\cite{Pasterski:2021raf}), they exhibit more freedom than what is needed for the transformation of NR waveforms.} It should be intuitively obvious that simple rotations satisfy these criteria, as do boosts. Given that the Poincaré group is the group of symmetries \emph{in} Minkowski space, we might also expect analogs of spacetime translations to be allowed, but it turns out that our use of the \emph{retarded} time $u$ and the fact that we are taking the $r\rightarrow\infty$ limit complicates matters.

To see this consider the following. A time translation $\delta t$ will surely affect only the retarded time via $u\mapsto u - \delta t$, but it is not obvious what to do with a space translation by some finite $\delta\vec{x}$. If we take $(\theta,\phi)$ as the direction of the unit vector $\hat{n}$, then the usual space translation changes the radial coordinate via $r\mapsto r+\delta\vec{x}\cdot\hat{n}$. Clearly this has no effect on the $r\rightarrow\infty$ limit, nor does it affect the angular coordinates \emph{in that limit}. However, it will affect the retarded time because of the mixture of time and space that the retarded time represents. Considering the prototypical retarded time $u\equiv t-r$, we can intuit the correct impact of a space translation, which is simply $u\mapsto u-\delta\vec{x}\cdot\hat{n}$. The slightly surprising feature here is that a space translation affects the retarded time in a \emph{direction-dependent} way. While a time translation has a monopolar effect, a space translation has a dipolar effect.

Naturally, this invites the intriguing question of whether higher-order multipoles might also be permissible. In fact, if we pose
\begin{align}
\label{eq:supertranslationonu}
u\mapsto u-\alpha(\theta,\phi),
\end{align}
with an arbitrary (sufficiently smooth) function $\alpha(\theta,\phi)$, we can check that the asymptotic form of the metric does not change under such a transformation. The function $\alpha(\theta,\phi)$ is exactly a supertranslation and is the ingredient that is needed for constructing the BMS group. It contains the usual spacetime translations as $\ell=0,1$ components, when viewed in terms of spherical harmonic modes, and proper supertranslations as $\ell\geq2$ components.

Like the usual spacetime translations, we may combine two
supertranslations by pointwise addition, and can thus turn the set of
supertranslations into an abelian group $\mathbb{T}$. In fact, this
abelian group $\mathbb{T}$ is a normal subgroup of the full BMS group,
with the factor group of the BMS group by $\mathbb{T}$ being the usual
restricted Lorentz group $\mathrm{SO}^{+}(3,1)$. The latter, however,
is not a normal subgroup. Therefore, the full BMS group is formally
the semidirect product $\mathbb{T}\rtimes\mathrm{SO}^{+}(3,1)$. Put
more simply, we can express any BMS transformation as a
supertranslation followed by a Lorentz transformation (see
Appendix~\ref{sec:BMSdecomposition} for more details).
For a boost with velocity $\vec{v}$, and thus a conformal factor\footnote{See Appendix D of Ref.~\cite{Boyle2016} for more on why boosts induce a conformal transformation of the celestial sphere.}
\begin{align}
\label{eq:conformalfactor}
k(\theta,\phi)\equiv\frac{\sqrt{1-|\vec{v}|^{2}}}{1-\vec{v}\cdot\hat{n}(\theta,\phi)},
\end{align}
where $\hat{n}(\theta,\phi)$ is the unit normal vector to the point $(\theta,\phi)$, the effect on the retarded time of a supertranslation $\alpha(\theta,\phi)$ followed
by this Lorentz transformation is simply Eq.~\eqref{eq:supertranslationonu} but with a Doppler factor, i.e.,
\begin{align}
u\mapsto k(\theta,\phi)(u-\alpha(\theta,\phi)).
\end{align}
Note that the $(\theta,\phi)$ angular coordinates are not affected by a supertranslation, but are altered through the more familiar rotation and relativistic aberration (see Ref.~\cite{Boyle2016}).

To summarize, the BMS group of future null infinity represents the remaining spacetime coordinate freedom that must be fixed when working with asymptotic data, like GWs, in Bondi gauge. It is made up of the usual Lorentz rotations and boosts, as well as a new set of transformations called supertranslations that extend the spacetime translations of the Poincaré group. While we hope that this section has provided some motivation for why the existence of supertranslations is perhaps expected from a mathematical perspective, in the next section we provide some physical intuition for why they are expected to be symmetries of future null infinity.

\subsection{Understanding supertranslations}
\label{sec:supertranslations}

For the moment, consider a proper supertranslation, i.e., a supertranslation which is not a time or space translation. In terms of spherical harmonics, this corresponds to a function $\alpha(\theta,\phi)$ whose spherical harmonic decomposition only consists of $\ell\geq2$ modes. Formally, supertranslations are rather elementary: they are angle-dependent offsets in the retarded time, as illustrated through Eq.~\eqref{eq:supertranslationonu}. Specifically, for each distant inertial observer at $\mathcal{I}^{+}$, a supertranslation corresponds to a simple change in the origin of the time coordinate. To understand why they are symmetries of future null infinity, we consider the following thought experiment.

Consider some asymptotically flat spacetime with an isolated astrophysical event---like a supernova explosion or a binary black hole merger---that is emitting radiation, such as photons or gravitational waves, outward in a spherical manner.  Furthermore, imagine a network of distant inertial observers that are surrounding this event, each at some finite radius from the event that need not be the same as the other observers. If these observers can communicate, then they could---in principle---use their knowledge of their locations relative to each other and to the central event to synchronize their clocks to ensure that their measurements of the central event are simultaneous in some sense. Now, consider what happens if these observers were placed at larger (but finite) radii. At these new, farther away positions, the signals that they use to synchronize their clocks will take longer to travel between them, but there is no fundamental obstacle to this synchronization \emph{in principle}.

However, as this network of inertial observers limits to an
\emph{infinite} radius away from this central event, then they become
causally disconnected from each other.
Formally, what this means is that each observer approaches some generator of $\mathcal{I}^{+}$, and because every generator of $\mathcal{I}^{+}$ is causally disconnected from any other, so too are the inertial observers. In this $r\rightarrow\infty$ limit, the observers can no longer synchronize their clocks, and therefore they can no longer ensure that they receive the same radiation from the central event at the same time. That is, the invariance with respect to standard time and space translations of the Poincaré group at finite radius yields, \emph{at infinite radius}, the invariance to the angle-dependent supertranslations. This fact, i.e., the notion that each and every point on $\mathcal{I}^{+}$ is causally disconnected from any other, is one way to intuit why supertranslations are indeed symmetries of future null infinity and thus elements of the BMS group.

\subsection{The effects of BMS transformations}
\label{sec:BMSeffects}

\figCyl

As mentioned earlier, waveforms are not invariant in any useful sense. At best, they are components of tensors defined with respect to a coordinate basis. Consequently, as we change the coordinates, the value of the waveform at each physical point will also change. Thus, when working with waveforms and their coordinate freedom we really have two main concerns: first, to transform the waveform, and second to transform the coordinates upon which the waveform is evaluated. As we have already explained, the latter can be expressed rather simply via the following.

Instead of working with Bondi coordinates, i.e., $(u,\theta,\phi)$, it tends to be simpler (at least mathematically) to use the complex stereographic coordinate
\begin{align}
\zeta\equiv e^{i\phi}\cot(\theta/2).
\end{align}
With this, the action of a BMS transformation on the coordinates of future null infinity can then be written as
\begin{align}
\label{eq:coordinatetransformation}
(u,\zeta)\mapsto\left(k(\zeta,\bar{\zeta})(u-\alpha(\zeta,\bar{\zeta})),\frac{a\zeta + b}{c\zeta + d}\right),
\end{align}
where the conformal factor $k(\zeta,\bar{\zeta})$ is
\begin{align}
k(\zeta,\bar{\zeta})\equiv\frac{1+|\zeta|^{2}}{|a\zeta+b|^{2}+|c\zeta+d|^2},
\end{align}
$(a,b,c,d)$ are complex coefficients with $ad-bc=1$ that encode the
Lorentz rotation and boost, and $\alpha(\zeta,\bar{\zeta})$ is a real,
smooth function that encodes the supertranslation.  See
App.~\ref{sec:BMSdecomposition} for details on how the M\"obius
transformation $(a,b,c,d)$ is related to usual Lorentz rotations
and boosts. As an illustration of the impact of these transformations,
see Fig.~\ref{fig:Cyl}, which shows how an example Lorentz boost,
space translation, and proper supertranslation change the retarded
time $u$ as a function of $\theta$.

From this coordinate transformation, by examining how the coordinate tetrad on $\mathcal{I}^{+}$ transforms, one can then ascertain how asymptotic data on $\mathcal{I}^{+}$ that describes the spacetime metric transforms. But what is this \emph{data}? Obviously, we want something on $\mathcal{I}^{+}$ that represents the gravitational wave strain $h$, since this is what our detectors measure. But is there other data that we should also be interested in, e.g., something describing the mass or the angular momentum of the spacetime that may be useful for, say, measuring the mass or the spin of an isolated black hole? While often neglected in the majority of the NR literature, the answer to this, as suggested by the isolated black hole example, is yes. 

Apart from the strain, to fully reconstruct the metric on $\mathcal{I}^{+}$ one also needs other information about the spacetime, which is neatly encoded in the five complex Weyl scalars $\Psi_{0,1,2,3,4}$, i.e., the components of the Weyl tensor~\cite{Newman:1961qr}.\footnote{Note that even though these are called ``scalars'', the Weyl scalars are also not invariant in any sense since they are still functions of the spacetime coordinates.} The Weyl tensor measures the curvature of spacetime. However, the individual Weyl scalars each have their own unique interpretation. Specifically, they can be viewed as
\begin{itemize}
	\item $\Psi_{0}$: ingoing radiation;
	\item $\Psi_{1}$: current multipole moment;
	\item $\Psi_{2}$: mass multipole moment;
	\item $\Psi_{3}$: news ($\sim\dot{h}$);
	\item $\Psi_{4}$: outgoing radiation ($\sim\ddot{h}$),
\end{itemize}
where dots represent time derivatives.\footnote{Note that other interpretations also exist, such as Ref.~\cite{Szekeres:1965ux}.} Consequently, it can be seen that $\Psi_{3}$ and $\Psi_{4}$ are actually degenerate with with the strain and thus only the strain and $\Psi_{0,1,2}$ are needed to measure physical features of the spacetime.\footnote{Formally, because of the Bianchi identities (see Eqs.~\eqref{eq:bianchiidentitites}), one only needs the strain on $\mathcal{I}^{+}$ and $\Psi_{0,1,2}$ on a certain time slice of $\mathcal{I}^{+}$ to fully reconstruct the asymptotic spacetime metric} As we will see later on in Sec.~\ref{sec:mathoverview}, $\Psi_{2}$ can be used to measure the mass of the spacetime and $\Psi_{1}$ can be used to measure the angular momentum of the spacetime.

If we now study how this asymptotic data transforms under a BMS transformation, one can ascertain through the tetrad transformation that the gravitational wave shear $\sigma$, which is related to the strain $h$ via $h\equiv2\bar{\sigma}$,\footnote{We introduce the shear because it makes many of the subsequent equations in this paper simpler; it is formally constructed in Eq.~\eqref{eq:straindefinition} by contracting complex dyads on the two-sphere with the angular part of the Bondi-Sachs metric.} and the Weyl scalars transform as~\cite{Bondi:1962px,Sachs:1962wk,NewmanUnti1962}
\begin{subequations}
\label{eq:datatransformation}
\begin{align}
\sigma'&=\frac{1}{k}e^{2i\lambda}\left[\sigma-\eth^{2}\alpha\right],\\
\Psi_{A}'&=\frac{1}{k^{3}}e^{(2-A)i\lambda}\sum\limits_{a=A}^{4}\begin{pmatrix}4-A\\a-A\end{pmatrix}\left(-\frac{1}{k}\eth u'\right)^{a-A}\Psi_{a}.
\end{align}
\end{subequations}
where $A\in\{0,1,2,3,4\}$, $\lambda$ is the ``spin phase''
\begin{align}
e^{i\lambda}\equiv\left[\frac{\partial\bar{\zeta}'}{\partial\bar{\zeta}}\left(\frac{\partial\zeta'}{\partial\zeta}\right)^{-1}\right]^{\frac{1}{2}}=\frac{c\zeta+d}{\bar{c}\bar{\zeta}+\bar{d}},
\end{align}
and $\eth$ and $\bar{\eth}$ are the usual GHP spin-weight operators~\cite{Geroch:1973am}. In spherical coordinates they can be written as
\begin{subequations}
\begin{align}
\eth f&=-\frac{1}{\sqrt{2}}\left(\sin\theta\right)^{+s}\left(\partial_{\theta}+i\csc\theta\partial_{\phi}\right)
\left[\left(\sin\theta\right)^{-s}f\right],\\
\bar{\eth} f&=-\frac{1}{\sqrt{2}}\left(\sin\theta\right)^{-s}\left(\partial_{\theta}-i\csc\theta\partial_{\phi}\right)
\left[\left(\sin\theta\right)^{+s}f\right].
\end{align}
\end{subequations}
When acting on spin-weighted spherical harmonics $\phantom{}_{s}Y_{(\ell,m)}$, they yield
\begin{subequations}
\begin{align}
\eth\left(\phantom{}_{s}Y_{(\ell,m)}\right)=+\frac{1}{\sqrt{2}}\sqrt{(\ell - s)(\ell + s + 1)}\phantom{}_{s+1}Y_{(\ell,m)},\\
\bar{\eth}\left(\phantom{}_{s}Y_{(\ell,m)}\right)=-\frac{1}{\sqrt{2}}\sqrt{(\ell + s)(\ell - s + 1)}\phantom{}_{s-1}Y_{(\ell,m)}.
\end{align}
\end{subequations}
Note that the conventions used here are consistent with the Moreschi-Boyle convention that is used across Refs.~\cite{Boyle2016,Mitman:2021xkq,Mitman:2022kwt,Iozzo:2021vnq,Moreschi:1988pc,Moreschi:1998mw,Dain:2000lij} and the code \texttt{scri}~\cite{scri,Boyle:2013nka,Boyle:2014ioa}.

With this information, one then has everything that is needed to
transform asymptotic data on $\mathcal{I}^{+}$ and therefore fix the
coordinate freedom of that data to match some canonical frame. This
notion of mapping asymptotic data to a certain frame is called BMS
frame fixing and will be reviewed in the context of NR simulations in Sec.~\ref{sec:BMSresults}.

\subsection{Pedagogical approach to memory}
\label{sec:memoryoverview}

Without delving into the complicated mathematics of Einstein's
equations, memory effects can be most easily understood as coming from
conservation laws that stem from the symmetries of null infinity: the
BMS group. Consequently, to provide some motivation behind why memory
effects exist in GR and how they can be studied, before examining them
with a more mathematical lens, as is performed in
Sec.~\ref{sec:mathoverview}, we will first provide some insight
by studying them with respect to BMS transformations.

With these additional symmetries of the BMS group, Noether's
theorem\footnote{%
  Note that Noether's theorem is modified for this situation since
  the nonzero flux of gravitational radiation implies that the charges
  are not conserved. We instead have
  \emph{non-}conservation laws precisely quantifying how much each charge changes.
  The interested reader can see Refs.~\cite{Dray:1984rfa,Wald:1999wa}
  to read about this subtlety.} %
interestingly implies that there should be a conservation law for
each supertranslation. Thus, because supertranslations are
effectively angle-dependent spacetime translations, one can easily
imagine that such a balance law might be of the form
\begin{align}
\label{eq:nearlypedagogicalconservationlaw}
0&=\,\text{``change in angle-dependent mass''}\nonumber\\
&\phantom{=.} +\,\text{``flux of angle-dependent energy''}
\end{align}
This expression, in fact, is nearly correct. The one piece of
information that is missing is that these two terms on the right-hand
side of Eq.~\eqref{eq:nearlypedagogicalconservationlaw} need not fully
cancel out. One way to realize this is by considering the scattering
of two particles in linearized gravity~\cite{Kovacs:1978eu}. First
note that, because of the linearization, there will be no energy
flux. However, because the particles scatter, there will still be a
change in the angle-dependent mass, i.e., a change in the mass
multipole moment. Then, because a change in the mass multipole moment
corresponds to a change in the strain---or, equivalently, the
shear---one can intuit that there should also be a term on the
left-hand side of Eq.~\eqref{eq:nearlypedagogicalconservationlaw} that
corresponds to the shear, e.g.,
\begin{align}
\label{eq:pedagogicalconservationlaw}
\text{``shear''}&=\,\text{``change in angle-dependent mass''}\nonumber\\
&\phantom{=.} +\,\text{``flux of angle-dependent energy''}
\end{align}
In fact, if one formally works through the mathematics, as is carried
out in Sec.~\ref{sec:mathoverview} and shown through
Eq.~\eqref{eq:supertranslationconservationlaw}, one discovers that the conservation
law stemming from supertranslations states exactly this.

\begin{figure*}[ht!]
	\includegraphics[width=\textwidth]{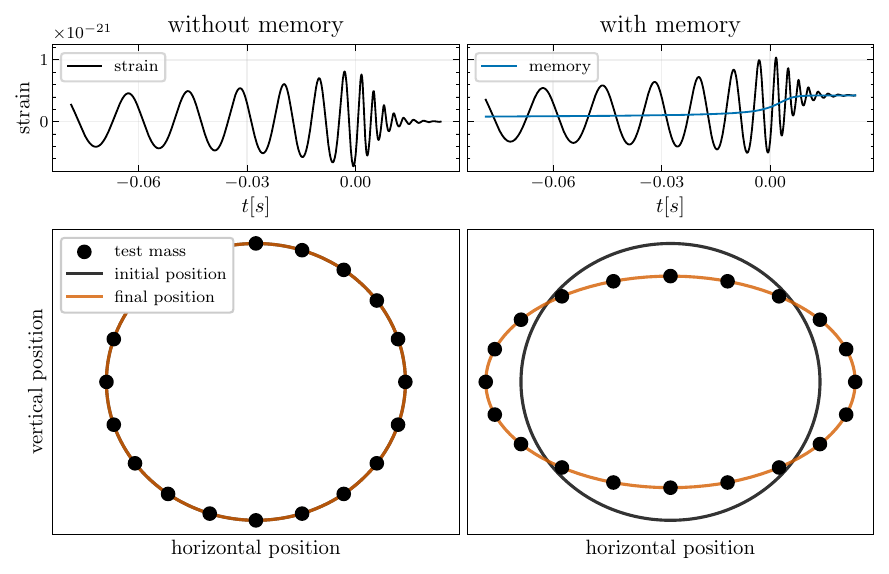}
	\caption{%
		\emph{Top}: Comparing the gravitational wave computed by a
		simulation of a binary black hole merger when memory effects are not included (left, incorrect) and when memory effects are
		included (right, correct). The binary black hole simulation is an equal-mass,
		aligned spin system with a total mass of $60\,M_{\odot}$, a
		luminosity distance of $400\,\mathrm{Mpc}$, an edge-on
		orientation, and equal dimensionless black hole spins of magnitude
		$0.6$ in the direction of the orbital angular momentum. Each
		waveform is shown in black and in the plot on the right, we show
		the contribution to the gravitational wave coming from the energy
		flux, i.e., a proxy for the memory, in blue. %
		\emph{Bottom}: The initial (black) and final (orange) positions of
		a series of test particles before and after the passage of a
		gravitational wave without (left) and with (right) memory
		traveling through the plane of the figure. Because of the
		orientation of the binary black hole system relative to the test
		particles, the GW exhibits a `+' polarization.  }
	\label{fig:MemoryObservers}
\end{figure*}

Using Eq.~\eqref{eq:pedagogicalconservationlaw}, one can then learn a
large amount of interesting physics about gravitational waves. For
one, we observe that, apart from the gravitational wave strain being
sourced by the mass multipole moment, there is also a contribution
from an angle-dependent energy flux. Furthermore, if we think about
the net change in the strain between non-radiative regimes, i.e., the
memory, we immediately find that there are two contributions: one from
the angle-dependent mass and one from the angle-dependent
energy. These unique contributions are the ordinary and null
contributions to the memory~\cite{Bieri:2013ada}.\footnote{%
  Previously, these contributions were called linear and nonlinear (or
  Christodoulou). However, they have since been renamed to better
  reflect the way in which they are sourced~\cite{Zeldovich:1974gvh,
    Braginsky:1985vlg, Braginsky:1987kwo, Christodoulou:1991cr,
    PhysRevD.45.520,Bieri:2013ada}.} %
They can be understood as being sourced by two types of physical
processes. In particular, ordinary memory is sourced by systems that contain
unbound masses, i.e., massive particles approaching timelike infinity~\cite{Bieri:2013ada,Madler:2016ggp,Zhang:2024tey}. Examples of systems that source this type of effect are hyperbolic black holes, neutron star disruptions, or even supernovae events. Meanwhile, null memory is
sourced by null radiation that escapes to future null infinity, e.g.,
gravitational or electromagnetic waves. For an example of what the null memory looks like in the GW produced by a binary black hole merger, see Fig.~\ref{fig:MemoryObservers}.

\section{Literature Review}
\label{sec:literaturereview}

Gravitational memory effects were first realized in 1974 when
Zel'dovich and Polnarev successfully calculated the gravitational
radiation that is produced by two objects on flyby, i.e., hyperbolic,
trajectories~\cite{Zeldovich:1974gvh}.\footnote{%
  Even earlier, in 1966, Newman and Penrose identified that, near null
  infinity, the strain of surfaces of constant retarded time will
  change between spacelike and future timelike
  infinity~\cite{Newman:1966ub}.} %
By working with Einstein's equations in linearized GR, they found
that, because the stress energy tensor exhibits a net change between
early and late times due to the change in the mass distribution of the
flyby objects, the strain will also exhibit such a net
change. Consequently, because this can also be understood as stemming
from a change in the Bondi mass aspect, the effect that they unearthed
is what we now call ordinary displacement memory. Later, in 1985, the
consequences of their result was elaborated upon by Braginski and
Grishchuk who first named this effect the ``memory
effect''~\cite{Braginsky:1985vlg}. Then, in 1987, Braginsky and Thorne
found a simple equation for the memory for scattering scenarios in
terms of the four-momentum of the ingoing and outgoing massive
particles~\cite{Braginsky:1987kwo}. It says that for a system of $N$
particles, the net change in the gravitational wave strain between
early and late times is
\begin{align}
\label{eq:linearmemory}
\Delta h_{ij}^{\mathrm{TT}}=\frac{4}{r}\Delta\sum\limits_{A=1}^{N}\frac{M_{A}}{\sqrt{1-v_{A}^{2}}}\left(\frac{v_{A}^{i}v_{A}^{j}}{1-v_{A}\cos(\theta_{A})}\right)^{\mathrm{TT}}
\end{align}
where $r$ is the distance from the observer to the source, $M_{A}$ is
the mass of particle $A$, $\vec{v}_{A}$ is the velocity with
$v_{A}^{i}$ the $i^{\mathrm{th}}$ component and $v_{A}$ the norm,
$\theta_{A}$ is the angle between $\vec{v}_{A}$ and the observer, and
the $\Delta$ in front of the sum on the right-hand side refers to the
difference in this sum evaluated for the outgoing and ingoing
particles.

After these early works, it was largely thought that memory effects
were understood. This opinion, however, was completely overturned when
in 1991 Christodoulou found that gravitational waves themselves will
also source a certain type of memory effect, through a subtle, but
detectable nonlinear interaction with
themselves~\cite{Christodoulou:1991cr}.\footnote{%
  This discovery was also realized by Payne (somewhat indirectly) as
  well as Blanchet and Damour in Refs.~\cite{Payne:1983rrr}
  and~\cite{Blanchet:1992br}.} %
Christodoulou obtained this important result by working with null
hypersurface equations and asymptotic limits to obtain an equation
that is similar in spirit to
Eq.~\eqref{eq:pedagogicalconservationlaw}. In particular, he found
that the strain is related to the flux of radiation through each point
on the two-sphere. Because of this connection to the energy flux, we
now call this effect the null displacement memory. A year later, in Ref.~\cite{PhysRevD.45.520} Thorne identified Christodoulou's finding as equivalent to that of
Ref.~\cite{Braginsky:1987kwo}, but with the various massive particles being
replaced by null radiation, i.e.,
\begin{align}
\label{eq:nonlinearmemory}
\Delta h_{ij}^{\mathrm{TT}}=\frac{4}{r}\int\frac{dE}{d\Omega'}\left(\frac{\xi^{i'}\xi^{j'}}{1-\cos(\theta')}\right)^{\mathrm{TT}}d\Omega',
\end{align}
where $E$ is the energy of the radiation, $\xi^{i'}$ is a unit vector
pointing from the source toward $d\Omega'$, and $\theta'$ is the angle
between $\xi^{i'}$ and the observer~\cite{PhysRevD.45.520}. In
Sec.~\ref{sec:mathoverview}, we will connect
Eqs.~\eqref{eq:linearmemory} and~\eqref{eq:nonlinearmemory} to the
modern interpretation of memory, which is more straightforward to
understand in terms of gravitational systems and radiation.

As for the BMS group, this was realized well before memory effects in
1962 by Bondi, Metzner, van der Burg, and
Sachs~\cite{Bondi:1962px,Sachs:1962wk}. However, the connection
between the BMS group and memory was not explicitly stated until
Refs.~\cite{Winicour:2014ska,Flanagan:2014kfa,Strominger:2014pwa} in
2014, even though this relationship has been largely understood
since, e.g., Refs.~\cite{Newman:1966ub, Geroch:1981ut,
  Ashtekar:1981bq, Dray:1984rfa, Ludvigsen:1989kg}. What makes this
history even more interesting, though, is in
Refs.~\cite{Strominger:2013jfa,Strominger:2014pwa} the duality between
memory effects and BMS symmetries was extended to be a triangle, since
it was found that BMS symmetries and memory effects could also be
related to soft theorems~\cite{Weinberg:1965nx}. This finding created
a large stir of interest in theory communities, since soft theorems
are inherently useful for studying the
$\mathcal{S}$-matrix of a quantum theory, so BMS symmetries seem integral to
understanding quantum gravity.
The ``soft limit'' means taking a particle's energy to zero,
so soft particles can not be directly measured by particle colliders.
However, memory effects
will soon be observed through GWs with current
detectors~\cite{Hubner:2019sly,Hubner:2021amk,Grant:2022bla}, and
therefore serve as
a natural probe of this exciting realm of physics. This excitement was
only further enhanced once more memory effects were unearthed through
this connection. In particular, in
Refs.~\cite{Cachazo:2014fwa,Pasterski:2015tva,Nichols:2018qac}, by
studying extensions of the BMS group to include
``superrotations''~\cite{deBoer:2003vf,Banks:2003vp,Barnich:2009se,Barnich:2010ojg},
i.e., extensions of the Lorentz transformations, two new memory
effects were found: the spin memory and the center-of-mass
memory. These two effects correspond to the net displacement that two
observers with an initial relative velocity will experience due to the
passage of transient radiation.\footnote{%
  They can also be viewed as the magnetic and electric components of the
  more general drift memory~\cite{Grant:2023ged}.} %
Ever since, the field of celestial holography, which aims to
establish a kind of holographic dictionary between gravitational scattering in
asymptotically flat spacetimes and a conformal field
theory on the celestial sphere, has reached unprecedented levels of
interest~\cite{Strominger:2017zoo, Pasterski:2019ceq,
  Raclariu:2021zjz, Pasterski:2021rjz, Pasterski:2021raf}.

At the same time as these theoretical developments regarding memory
were occurring, significant progress in resolving memory effects in
numerical simulations was also being achieved. In particular, in early
2010 Ref.~\cite{Pollney:2010hs} was able to successfully simulate the
memory sourced by a binary black hole merger using a more robust
method for extracting the NR waveforms at future null infinity:
Cauchy-characteristic evolution (CCE).\footnote{%
  We will elaborate more on this waveform extraction procedure, as
  well as others, in Sec.~\ref{sec:codeframeworks}.} %
Later, in 2020, Ref.~\cite{Mitman:2020pbt} performed a similar series
of simulations using a more efficient version of the code and
calculated the individual contributions to the gravitational wave
strain in terms of the charges and fluxes of
Eqs.~\eqref{eq:pedagogicalconservationlaw}
and~\eqref{eq:conservationlaws}. This showed that, as expected, the
memory in binary black hole mergers is indeed sourced by the null
memory.

One complexity that arose, however, was, with these new NR waveforms
that contained memory, it was not exactly obvious how to compare these
finite waveforms to post-Newtonian (PN) waveforms that had information
about the entire past history of the binary inspiral and thus had a
larger prediction for the instantaneous value of the
strain~\cite{Wiseman:1991ss}. This was important because if one wanted
to construct a hybridization between these NR waveforms and PN
waveforms to build a waveform model from, the hybridization would fail
because the two predictions would not agree (see, e.g.,
Fig.~\ref{fig:BMSFrameIssue}). This was resolved in a series of works which
established a program now called the BMS frame fixing
program~\cite{Mitman:2021xkq, MaganaZertuche:2021syq,
  Mitman:2022kwt}. Ever since, numerous findings and advancements in
gravitational wave physics using these developments have been
made~\cite{Mitman:2022qdl, Yoo:2023spi, Grant:2023jhd, Zhu:2023fnf,
  Zhu:2024rej}.

\section{Mathematical Overview}
\label{sec:mathoverview}

Recall that Eq.~\eqref{eq:BSmetric} provides the general form of the metric in Bondi gauge, in terms of the functions $U$, $\beta$, $\mathcal{U}^{A}$, and $\gamma_{AB}$, where $A$ and $B$ range over $(\theta, \phi)$. In general, these functions can each depend on all four coordinates. However, Bondi gauge demands specific behavior in the limit of $r \to \infty$.  To proceed, we expand this behavior in powers of $1/r$:
\begin{subequations}
\label{eq:falloffconditions}
\begin{align}
U&=1-\frac{2m}{r}-\frac{2\mathcal{M}}{r^{2}}+\mathcal{O}(r^{-3}),\\
\beta&=\frac{\beta_{0}}{r}+\frac{\beta_{1}}{r^{2}}+\frac{\beta_{2}}{r^3}+\mathcal{O}(r^{-4}),\\
\mathcal{U}^{A}&=\frac{U^{A}}{r^{2}}+\frac{1}{r^{3}}\Big[-\frac{2}{3}N^{A}+\frac{1}{16}D^{A}\left(C_{BC}C^{BC}\right)\nonumber\\
&\phantom{=.\frac{U^{A}}{r^{2}}+\frac{1}{r^{3}}\Big[}+\frac{1}{2}C^{AB}D^{C}C_{BC}\Big]+\mathcal{O}(r^{-4}),\\
\label{eq:angularpartexpansion}
\gamma_{AB}&=q_{AB}+\frac{C_{AB}}{r}+\frac{D_{AB}}{r^{2}}+\frac{E_{AB}}{r^{3}}+\mathcal{O}(r^{-4}),
\end{align}
\end{subequations}
where the various coefficients on the right-hand sides are functions
of $(u,\theta^{A})$ only, and $q_{AB}(\theta^{A})$ is the metric on
the two-sphere, i.e., in the usual spherical coordinates
$q_{AB}(\theta,\phi)dx^{A}dx^{B}=d\theta^{2}+\sin^{2}\theta\,d\phi^{2}$. Of
these functions, the most important ones are the Bondi mass
aspect $m$, the Bondi angular momentum aspect $N^{A}$, and finally the
shear tensor $C_{AB}$, whose retarded time derivative is the Bondi
news tensor $N_{AB}\equiv\partial_{u}C_{AB}$, which characterizes the
presence of radiation in a spacetime.

At this point one can then impose Einstein's equations to obtain evolution
equations for the various functions. Specifically, from the
$\mathcal{O}(uu,r^{-2})$ and $\mathcal{O}(uA,r^{-2})$ parts\footnote{%
  Here the first argument corresponds to the component of the metric
  tensor that is being examined, while the second argument corresponds
  to the relevant term in the $1/r$ expansion.} %
of the Einstein tensor one obtains~\cite{Flanagan:2015pxa}
\begin{subequations}
\label{eq:evolution}
\begin{align}
\partial_{u}m&=\frac{1}{4}D_{A}D_{B}N^{AB}-\frac{1}{8}N_{AB}N^{AB},\\
\partial_{u}\hat{N}_{A}&=\frac{1}{4}\left(D_{B}D_{A}D_{C}C^{BC}-D^{2}D^{B}C_{AB}\right)\nonumber\\
&\phantom{=.}-\Big[\frac{3}{8}[\left(N_{AB}D_{C}C^{BC}-C_{AB}D_{C}N^{BC}\right)\nonumber\\
&\phantom{=.-\Big[}-\frac{1}{8}\left(N^{BC}D_{B}C_{AC}-C^{BC}D_{B}N_{AC}\right)\Big]\nonumber\\
&\phantom{=.}-uD_{A}\partial_{u}m,
\end{align}
\end{subequations}
where $\hat{N}^{A}$ is the angular momentum aspect with the Wald-Zoupas correction~\cite{Wald:1999wa}:
\begin{align}
  \hat{N}_{A}&\equiv N_{A}-uD_{A}m\nonumber\\
  &\phantom{=.}-\frac{1}{4}C_{AB}D_{C}C^{BC}-\frac{1}{16}D_{A}\left(C_{BC}C^{BC}\right).
\end{align}

By contracting these evolution equations with dyads on the two-sphere\footnote{See, e.g., Ref.~\cite{Mitman:2020pbt} for more details.}
and making use of the Bianchi identities for the Weyl scalars, i.e.,
\begin{subequations}
	\label{eq:bianchiidentitites}
	\begin{align}
	\dot{\Psi}_{0}&=\eth\Psi_{1}+3\sigma\Psi_{2},\\
	\dot{\Psi}_{1}&=\eth\Psi_{2}+2\sigma\Psi_{3},\\
	\label{eq:psi2bianchiidentity}
	\dot{\Psi}_{2}&=\eth\Psi_{3}+\sigma\Psi_{4},\\
	\label{eq:impsi2bianchiidentity}
	\mathrm{Im}\left[\Psi_{2}\right]&=-\mathrm{Im}\left[\eth^{2}\bar{\sigma}+\sigma\dot{\bar{\sigma}}\right],\\
	\Psi_{3}&=-\eth\dot{\bar{\sigma}},\\
	\Psi_{4}&=-\ddot{\bar{\sigma}},
	\end{align}
\end{subequations}
one can
then rewrite Eqs.~\eqref{eq:evolution} rather simply in terms of spin-weighted
functions as
\begin{subequations}
\label{eq:evolutionSW}
\begin{align}
  \label{eq:massaspectevolution}
  \partial_{u}m&=-\partial_{u}\mathrm{Re}\left[\Psi_{2}+\sigma\dot{\bar{\sigma}}\right],\\
  \label{eq:angularmomentumaspectevolution}
  \partial_{u}\hat{N}&=-\partial_{u}\left[\Psi_{1}+\sigma\eth\bar{\sigma}+u\eth m+\frac{1}{2}\eth\left(\sigma\bar{\sigma}\right)\right].
\end{align}
\end{subequations}
Then, by reorganizing terms (for Eq.~\eqref{eq:supertranslationconservationlaw}) or following the derivation in Ref.~\cite{Mitman:2020pbt} (for Eq.~\eqref{eq:otherconservationlaw}), one can rewrite Eqs.~\eqref{eq:evolutionSW} with the shear on the left-hand side as
\begin{subequations}
\label{eq:conservationlaws}
\begin{align}
\label{eq:supertranslationconservationlaw}
\mathrm{Re}\left[\eth^{2}\bar{\sigma}\right]&=m+\mathcal{E},\\
\label{eq:otherconservationlaw}
\mathrm{Im}\left[\eth^{2}\bar{\sigma}\right]&=\partial_{u}\mathrm{Im}\left[\eth^{-1}\left(\hat{N}+\mathcal{J}\right)\right],
\end{align}
\end{subequations}
where
\begin{subequations}
\label{eq:fluxes}
\begin{align}
\label{eq:energyflux}
\mathcal{E}&\equiv\int_{-\infty}^{u}|\dot{\sigma}|^{2}\,du,\\
\label{eq:angularmomentumflux}
\mathcal{J}&\equiv\int_{-\infty}^{u}\frac{1}{2}\left(3\dot{\sigma}\eth\bar{\sigma}-3\sigma\eth\dot{\bar{\sigma}}+\bar{\sigma}\eth\dot{\sigma}-\dot{\bar{\sigma}}\eth{\sigma}\right)\,du.
\end{align}
\end{subequations}
In Eqs.~\eqref{eq:fluxes}, $\mathcal{E}$ can be thought of as the total energy flux measured at $\mathcal{I}^{+}$ and $\mathcal{J}$ as the angular momentum flux. Eq.~\eqref{eq:supertranslationconservationlaw} is called the
supertranslation conservation law and can be thought of as the
conservation law that stems from the supertranslation symmetry of
$\mathcal{I}^{+}$; Eq.~\eqref{eq:otherconservationlaw}, however, is
not often presented in the literature, in part because it does not
have as clear of an interpretation as
Eq.~\eqref{eq:supertranslationconservationlaw} as being the
conservation law corresponding to a symmetry contained in the BMS
group. Nonetheless, other symmetry groups of future null infinity have
also been proposed that contain more symmetries than the BMS group for
which Eq.~\eqref{eq:otherconservationlaw} can be understood as a
conservation
law~\cite{deBoer:2003vf,Banks:2003vp,Barnich:2009se,Barnich:2010ojg}. Specifically,
Eq.~\eqref{eq:otherconservationlaw} has been referred to as the
superrotation conservation law. These other groups are extensions of
the BMS group. They are obtained by relaxing the fall off conditions
in Eqs.~\eqref{eq:falloffconditions} to be less restrictive, which
ends up enabling the existence of other symmetries that extend the
usual Lorentz transformations. But, because in this review we restrict
our attention to focus on the BMS group, for the remainder of this
paper we will not consider these other symmetry groups and instead
refer the interested reader to explore these extensions further in
Refs.~\cite{deBoer:2003vf,Banks:2003vp,Barnich:2009se,Barnich:2010ojg}.

To provide more motivation as to why, at least to some degree,
Eqs.~\eqref{eq:conservationlaws} should be viewed as conservation
laws, let us first consider what the Bondi mass aspect and the angular
momentum aspect represent.\footnote{Note that the word aspect here corresponds to the fact that the mass and angular momentum aspects can be thought of as angle-dependent generalizations of mass and angular momentum.} In particular, from these aspects one can
construct Poincaré charges, i.e., the translation, rotation, and boost
charges~\cite{GomezLopez:2017kcw,Dray:1984rfa,Dray:1984gz,Streubel1978}
\begin{subequations}
\label{eq:Poincarecharges}
\begin{align}
\label{eq:momentumcharge}
P^{a}(u)&=\frac{1}{4\pi}\int_{S^{2}}n^{a}m\,d\Omega,\\
\label{eq:angularmomentumcharge}
J^{a}(u)&=\frac{1}{4\pi}\int_{S^{2}}\text{Im}\left[\left(\bar{\eth}n^{a}\right)\hat{N}\right]\,d\Omega,\\
\label{eq:boostcharge}
K^{a}(u)&=\frac{1}{4\pi}\int_{S^{2}}\text{Re}\left[\left(\bar{\eth}n^{a}\right)\hat{N}\right]\,d\Omega,
\end{align}
\end{subequations}
where $n^{a}$ with $a\in{t,x,y,z}$ is the four vector
\begin{subequations}
\begin{align}
n^{t}&=1\nonumber\\
&=\sqrt{4\pi}Y_{(0,0)},\\
n^{x}&=\sin\theta\cos\phi\nonumber\\
&=\sqrt{\frac{4\pi}{3}}\left[\frac{1}{\sqrt{2}}\left(Y_{(1,-1)}-Y_{(1,+1)}\right)\right],\\
n^{y}&=\sin\theta\sin\phi\\
&=\sqrt{\frac{4\pi}{3}}\left[\frac{i}{\sqrt{2}}\left(Y_{(1,-1)}+Y_{(1,+1)}\right)\right],\\
n^{z}&=\cos\theta\\
&=\sqrt{\frac{4\pi}{3}}Y_{(1,0)},
\end{align}
\end{subequations}
with each component being a spin-weight 0 function. Meanwhile, the quantities $\mathcal{E}$ and $\mathcal{J}$ defined in
Eqs.~\eqref{eq:fluxes} are the usual energy and angular momentum
fluxes. So, if one takes Eqs.~\eqref{eq:conservationlaws} and instead
writes them in terms of spherical harmonics, it can readily be seen
that, since $\eth^{2}\bar{\sigma}$ has no $\ell=0$ or $\ell=1$
components, these equations are the four and angular momentum
conservation laws that correspond to the translation and rotation
symmetries. Obtaining such charges can be achieved by following the
prescription of Wald and Zoupas~\cite{Wald:1999wa} or the derivation
of these exact charges by Dray and Streubel~\cite{Dray:1984rfa}.

As for the $\ell\geq2$ components of Eqs.~\eqref{eq:conservationlaws},
these parts correspond to the supertranslation (and superrotation)
conservation laws mentioned earlier. This can be seen, for example,
from the fact that here the energy flux that appears in
Eq.~\eqref{eq:supertranslationconservationlaw} is a function of
angular coordinates and thus corresponds to the energy radiated at
each point on the celestial two-sphere, which is reminiscent of the
pedagogical example presented in
Sec.~\ref{sec:memoryoverview}. However, to view
Eq.~\eqref{eq:supertranslationconservationlaw} as a supertranslation
conservation law, we should also identify the charge that corresponds
to supertranslations. Naively, one might expect that the
supertranslation charge is simply Eq.~\eqref{eq:momentumcharge}, but
with $n^{a}$ replaced by some arbitrary function on the two-sphere,
e.g.,
\begin{align}
\label{eq:supermomentumcharge}
\mathcal{P}_{(\ell,m)}(u)=\frac{1}{4\pi}\int_{S^{2}}\left[\sum\limits_{\ell\geq0,|m|\leq\ell}\alpha_{(\ell,m)}Y_{(\ell,m)}\right]m\,d\Omega,
\end{align}
where $\alpha_{(\ell,m)}$ are spherical harmonic coefficients for some
real-valued, smooth function $\alpha(\theta,\phi)$. This, in fact, is
a reasonable hypothesis for this charge. In particular, the only
possible expressions for the supertranslation charge, or what some
call the ``supermomentum'', are
\begin{align}
\label{eq:generalsupermomentum}
\mathcal{P}_{p,q}(u,\theta,\phi)=\Psi_{2}+\sigma\dot{\bar{\sigma}}+p\left(\eth^{2}\bar{\sigma}\right)-q\left(\bar{\eth}^{2}\sigma\right),
\end{align}
where $p$ and $q$ are real numbers~\cite{Dray:1984rfa}. This was
pointed out by Dray and Streubel in Ref.~\cite{Dray:1984rfa} and was
also later independently realized by Wald and Zoupas in
Ref.~\cite{Wald:1999wa}. From this supermomentum expression, it can be
easily shown that if $p+q=1$ then the supermomentum is real and if
$p=q$ there is no supermomentum flux in Minkowski space, both of which
are nice properties~\cite{Dray:1984rfa}. Consequently, the natural
choice of supermomentum is the Geroch (G) supermomentum with
$p=q=\frac{1}{2}$~\cite{Geroch1977,Dray:1984rfa}
\begin{align}
\label{eq:Gerochsupermomentum}
\mathcal{P}_{G}(u,\theta,\phi)&\equiv\Psi_{2}+\sigma\dot{\bar{\sigma}}+i\mathrm{Im}\left[\eth^{2}\bar{\sigma}\right]\nonumber\\
&=-m,
\end{align}
which, up to a sign, is exactly the Bondi mass aspect. Thus, the naive
guess for the supermomentum being that which appears in
Eq.~\eqref{eq:supermomentumcharge} is indeed correct.

At this point, having established the BMS charges, i.e.,
Eqs.~\eqref{eq:Poincarecharges} as well as
Eq.~\eqref{eq:supermomentumcharge}, we can now return to the
examination of Eqs.~\eqref{eq:conservationlaws} as the BMS
conservation laws. In particular, since $\eth^{2}\bar{\sigma}$ has no
$\ell=0$ or $\ell=1$ components when written in terms of spherical
harmonics, we readily find that the $\ell=0$ component of
Eq.~\eqref{eq:supertranslationconservationlaw} is stating energy
conservation, while the $\ell=1$ component is stating linear momentum
conservation. As for Eq.~\eqref{eq:otherconservationlaw}, the $\ell=0$
component of this equation is trivially zero. But, the $\ell=1$ is
stating angular momentum conservation. And, finally, if we now examine
the $\ell\geq2$ components of
Eq.~\eqref{eq:supertranslationconservationlaw}, we readily find a
statement regarding the conservation of supermomentum. For
Eq.~\eqref{eq:otherconservationlaw}, this has sometimes been viewed as
a statement regarding the conservation of super angular momentum, but
we stress that to do so requires extending the symmetry group of
future null infinity to be larger than the BMS group~\cite{Compere:2019gft}.

Now, besides providing a clear and straightforward connection between
the various BMS symmetries and the conservation of charges and fluxes
at null infinity, Eqs.~\eqref{eq:conservationlaws} also provide a
unique and useful means for studying gravitational waves. In
particular, because the right-hand side of
Eqs.~\eqref{eq:conservationlaws} contain the shear, which is related
to the gravitational wave strain $h$ via $h=2\bar{\sigma}$, one can
readily use Eqs.~\eqref{eq:conservationlaws} to study contributions to
the strain in terms of BMS charges and fluxes. In particular, as
explained in Sec.~\ref{sec:memoryoverview}, the contribution from
the charges in Eqs.~\eqref{eq:conservationlaws} correspond to the
ordinary memory while the flux contributions corresponds to the null
memory.

\section{Numerical Code Frameworks}
\label{sec:codeframeworks}

In this section we briefly outline the code frameworks that are
required to simulate a binary black hole merger and extract the
waveform data to future null infinity. Readers who are primarily
interested in examining the waveforms output by the numerical
simulations and how they can be studied using
Eqs.~\eqref{eq:conservationlaws}, rather than the details of the
numerical simulation, may skip to Sec.~\ref{sec:memoryresults}.

When simulating a binary black hole merger, there are two types of
numerical evolutions that need to be run.\footnote{%
  In practice constructing initial data for the binary that one aims
  to simulate is also an important part of the simulation. However,
  because here we aim to review the parts of the simulation that
  produce meaningful waveform data, we will restrict ourselves to a
  discussion of the different types of evolutions. The interested
  reader can look to Ref.~\cite{Cook:2000vr} for more on initial data
  construction.} %
The first, and the bulk of the computation, is what is typically
called the Cauchy evolution. This part of the NR simulation involves
solving Einstein's equations on a finite region of spacetime near the
binary black holes to obtain the metric and its derivatives at a
finite radius.

Once the Cauchy evolution is complete and the metric and its
derivatives are obtained at a finite radius, it is then possible to
run a waveform extraction to obtain the waveform data at future null
infinity. This is what we use as a proxy for the data that GW
detectors should observe on Earth, because the extra information from
being at a finite distance from the GW-sourcing event are higher order
in the $1/r$ expansion of the angular part of the metric and should
therefore be highly subdominant. Ultimately, there are two ways this
can be performed:
\begin{itemize}
  \item extrapolation, which consists of fitting metric data ($\gamma_{AB}$ in Eq.~\eqref{eq:angularpartexpansion}) to polynomials in $1/r$ to extract waveform data at future null infinity~\cite{Boyle:2019kee}; or
  \item a characteristic evolution, which consists of solving Einstein's
    equations on hypersurfaces that connect a finite radius
    worldtube to future null infinity.
\end{itemize}
The primary output of these extraction methods is the gravitational wave strain $h$, which, with respect to the shear tensor $C_{AB}$ in Eq.~\eqref{eq:angularpartexpansion}, is defined via
\begin{align}
\label{eq:straindefinition}
h(u,\theta,\phi)&\equiv\bar{q}_{A}\bar{q}_{B}C_{AB}\nonumber\\
&=\sum\limits_{\ell\geq2,|m|\leq\ell}h_{(\ell,m)}(u)\phantom{}_{-2}Y_{(\ell,m)}(\theta,\phi)
\end{align}
and is a spin-weight $-2$ function often decomposed into spin-weight $-2$ spherical harmonics $\phantom{}_{-2}Y_{(\ell,m)}$ with the complex coefficients $h_{(\ell,m)}(u)$. In Eq.~\eqref{eq:straindefinition}, 
$q_{A}$ is the dyad $q_{A}\equiv-\left(1,i\sin(\theta)\right)/\sqrt{2}$ (see, e.g., Ref.~\cite{Mitman:2020pbt}).

As one may imagine, the characteristic evolution, albeit more
challenging to perform, is much more accurate than extrapolation: both
in terms of the numerical precision that can be achieved as well as
ensuring that the expected physics is accurately captured.\footnote{%
  In particular, waveform data produced using extrapolation is known
  to not capture memory effects. We will discuss this failure, and how
  it can be mitigated, more in Sec.~\ref{sec:memorycorrection}.} %
Effectively, the way that the characteristic evolution works, which is
formally called a Cauchy-characteristic evolution (CCE), is through
the following. First, treat the worldtube at some finite radius and
the initial null hypersurface that connects the worldtube to future
null infinity as two sets of initial data. Then, reduce Einstein's
equations to a series of ordinary differential equations
(ODEs). Finally, solve this series of ODEs by integrating in retarded
time to obtain data for the waveforms on subsequent
null hypersurfaces and eventually the whole of future null
infinity. By performing this sequence of tasks, one can then
accurately compute the strain as well as the Weyl scalars at future null
infinity. One subtlety, however, is that choosing the initial data on
the first null hypersurface is a highly nontrivial problem. An
incorrect choice of this initial data effectively amounts to putting
the output waveform data in some arbitrary BMS frame that needs to be
manipulated in order to perform robust waveform model comparisons or
analyses. This issue is covered in more detail in
Sec.~\ref{sec:BMSintro}.

The idea of CCE was first theorized in 1996~\cite{Bishop:1996gt,Bishop:1997ik}. However, it was not until 2009 when a CCE code was first used on a binary black hole simulation~\cite{Reisswig:2009rx,Reisswig:2009us,Babiuc:2010ze,Pollney:2010hs}. This version of CCE was run using the finite-difference Pitt Null code~\cite{Reisswig:2009rx,Reisswig:2009us,Babiuc:2010ze,Pollney:2010hs}. Later, in 2014, an improved version of CCE using spectral methods was incorporated into the SpEC code~\cite{Handmer:2014qha,Handmer:2015dsa,Handmer:2016mls}. And, finally, throughout 2020 and 2021 an even more-improved version of CCE that could extract the Weyl scalars was developed by Ref.~\cite{Moxon:2020gha} and incorporated into the SpECTRE code~\cite{spectrecode} by Ref.~\cite{Moxon:2021gbv}. This version of CCE is the most advanced version and is what will be used throughout this review.

\section{Numerical Waveforms \\and Memory Effects}
\label{sec:memoryresults}

\begin{figure*}[ht!]
  \includegraphics[width=\textwidth]{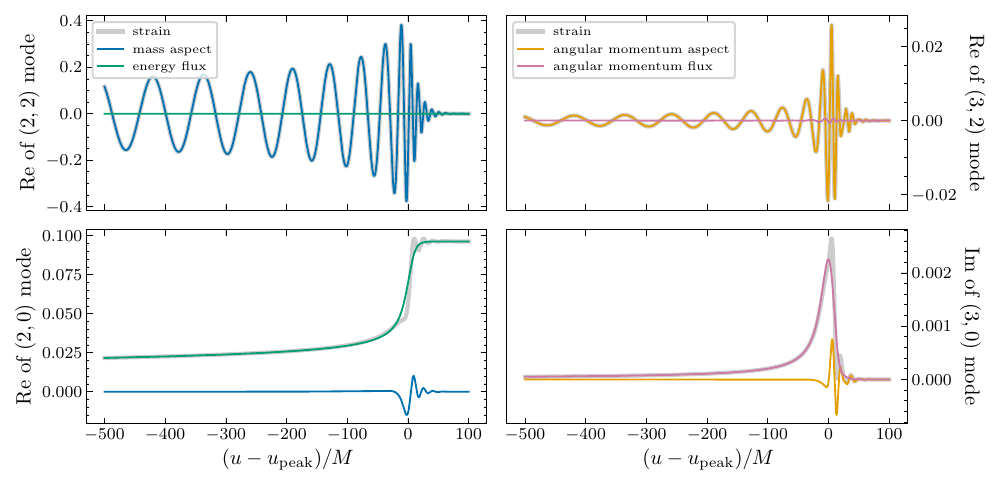}
  \caption{%
    Four different spin-weight $-2$ spherical harmonic modes of the
    gravitational wave strain and the contributions from the BMS
    charges and fluxes appearing in the right-hand side of
    Eqs.~\eqref{eq:conservationlaws}. The system is a mass ratio
    $q=1.22$ binary black hole, whose black hole spins are parallel to
    the system's orbital angular momentum and have the values
    $\chi_{z}^{(1)}=0.33$ and $\chi_{z}^{(2)}=-0.44$. Note that this
    system (SXS:BBH:0305) has parameters which are consistent with those of the first GW detection GW150914.
    \emph{Top Left:} The real part of the $(\ell,m)=(2,2)$ mode of the
    strain. The Bondi mass aspect's and energy flux's contributions
    are shown in blue and green. %
    \emph{Bottom Left:} Identical to that shown in the top left panel,
    but for the real part of the $(2,0)$ mode. %
    \emph{Top Right:} The real part of the $(3,2)$ mode of the
    strain. The angular momentum aspect's and flux's contributions are
    shown in orange and purple. \emph{Bottom Right:} Identical to that
    shown in the top right panel, but for the imaginary part of the
    $(3,0)$ mode. The horizontal axis for each plot is the retarded
    time $u$, with $u_{\mathrm{peak}}$ the peak of the $L^{2}$ norm of
    the news over the two sphere.}
  \label{fig:ChargeFlux}
\end{figure*}

\subsection{Examining conservation laws in Eqs.~\eqref{eq:conservationlaws}}
\label{sec:conservationlawresults}

With the pedagogical and mathematical theory behind the BMS group, its
corresponding conservation laws, and memory effects presented in
earlier sections, we now turn to a numerical study
of GWs in the context of asymptotics. In particular,
to highlight the usefulness of Eqs.~\eqref{eq:conservationlaws}, we
first present Fig.~\ref{fig:ChargeFlux} to illustrate how the gravitational
wave strain can be easily interpreted in terms of BMS charges and
fluxes. In this figure we show four spin weight $-2$ spherical
harmonic modes of the gravitational wave strain as well as the
contributions to the strain coming from the BMS charges and fluxes in
the right-hand sides of Eqs.~\eqref{eq:conservationlaws}. The system
considered is a mass ratio $q=1.22$ binary black hole, whose black
hole spins are parallel to the system's orbital angular momentum and
have the dimensionless magnitudes $\chi_{z}^{(1)}=0.33$ and
$\chi_{z}^{(2)}=-0.44$. These parameters resemble the most likely
parameters of the first gravitational wave detection, GW150914~\cite{LIGOScientific:2016aoc}.

Let us first consider what is shown in the top left panel, i.e., the
$(\ell,m)=(2,2)$ mode. As can be seen, the strain is nearly entirely
sourced by the Bondi mass aspect. This, however, is expected because
the Bondi mass aspect can effectively be thought of as the mass
multipole moment. So all that this plot is showing is an illustration
of the fact that gravitational waves are predominantly sourced by the
mass quadrupole moment. If we now examine the top right panel, we see
a similar phenomenon, but now illustrating the fact that gravitational
waves are also sourced by the current multipole moment. In this panel
we show the $(3,2)$ mode of the strain as well as the two
contributions from the angular momentum aspect and the angular
momentum flux. Like the top left panel, we find that the strain is
primarily sourced by the charge in
Eqs.~\eqref{eq:conservationlaws}. And, since the angular momentum
aspect can be related to the current multipole moment, this plot also
highlights the fact that the current multipole moment also sources the
strain, albeit subdominantly.

The perhaps more interesting panel in Fig.~\ref{fig:ChargeFlux}, however, is
the bottom left panel. In this plot we now show the $(2,0)$ mode of
the strain as well as the contributions from both the Bondi mass aspect and
the energy flux. As can be seen, instead of the strain being
sourced by the Bondi mass aspect, we instead find that there is a
large and dominating contribution from the energy flux. Furthermore,
there is the unique phenomenon that the strain no longer decays to
zero; this is the memory effect!\footnote{%
  One also may observe that there is a nonnegligible contribution from
  the Bondi mass aspect near the peak of the strain. This is because
  when the black holes are merging, they form an excited remnant black
  hole that emits gravitational waves in a process called the
  ringdown. The contribution from the mass aspect is exactly this
  ringing that occurs as the remnant black hole settles to be in a
  state of equilibrium, i.e., a Kerr black hole} %
More specifically, this is the term that we associate with the
memory. What we mean by this is the following. Formally, gravitational
wave memory is a phenomenon that can only be measured between two
non-radiative regions of spacetime, e.g., before and after the passage
of a burst of gravitational radiation. In practice, however,
gravitational wave detectors are not really freely-falling\footnote{%
  This is because the test masses in a gravitational wave detector are
  influenced by actuators that are always working to restore the
  initial configuration of the detector.} %
and analyze the strain in the frequency domain. Thus, to measure
memory we need to associate the memory with some nonzero frequency,
which is most naturally whatever the frequency of the energy flux is,
because it is the source of the memory for these binary systems, as
illustrated through the bottom left panel of Fig.~\ref{fig:ChargeFlux}. So,
for the remainder of the paper when we refer to memory what we will
really be referring to is the evolution of the memory, as measured
through the BMS fluxes.

Finally, in the bottom right panel of Fig.~\ref{fig:ChargeFlux} we again
highlight a similar result to that of the bottom left panel, but now
illustrated through the imaginary part of the $(3,0)$ mode of the
strain. As is shown, the strain is again predominantly sourced by the
flux contribution, but now the contribution looks like a delta
function, rather than a step function; this is the spin memory
effect. As outlined in Sec.~\ref{sec:literaturereview}, unlike the
displacement memory, which affects initially co-moving observers, the
spin memory instead affects observers with a non-zero initial relative
velocity. This can be understood in part by thinking about the time
integral of the strain, in which case this contribution would instead
manifest as a step function.

\subsection{Detectability of Memory Effects}
\label{sec:detectability}

\begin{figure}[t]
  \includegraphics[width=0.5\textwidth]{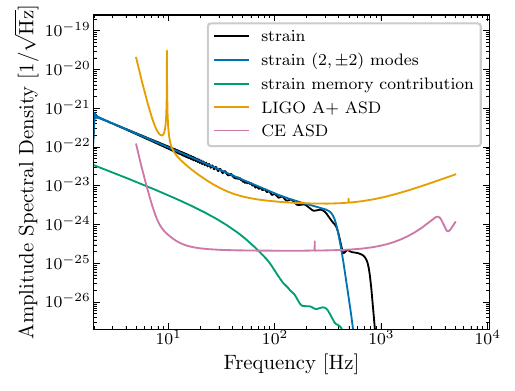}
  \caption{%
    Amplitude spectral density of the GW strain (black) evaluated at
    $(\theta,\phi)=(\pi/2,0)$ and the contributions to the strain from
    the $(2,\pm 2)$ modes (blue) and energy flux (green). The NR
    system is an equal-mass, aligned spin binary with a total mass of
    $60\,M_{\odot}$, a luminosity distance of $400\,\mathrm{Mpc}$, and
    equal dimensionless black hole spins of magnitude $0.6$ in the
    direction of the orbital angular momentum. This simulation is used
    to show a more optimistic observation of the memory. As a
    reference, we include the LIGO A+ noise curve in orange and the CE
    noise curve in magenta.}
  \label{fig:MemoryFreq}
\end{figure}

While Fig.~\ref{fig:ChargeFlux} is useful in that it clearly illustrates the
fact that the energy flux is the contribution responsible for the memory
exhibited by the gravitational wave strain, it does not provide the
overall magnitude of this effect. This is because when the strain is
evaluated at a point on the celestial sphere, e.g., if the strain were
observed by a gravitational wave detector, one needs to consider the
sum of the strain's modes weighted by the spin weighted spherical
harmonics. As a result, while the memory looks fairly prominent in the
bottom left panel of Fig.~\ref{fig:ChargeFlux}, if one evaluates the strain
at a point on the sky the memory can be noticeably suppressed. This is
in part because the $(2,0)$ spin weight $-2$ spherical harmonic is
\begin{align}
\phantom{}_{-2}Y_{(2,0)}(\theta,\phi)=\frac{1}{4}\sqrt{\frac{15}{2\pi}}\sin(\theta)^{2},
\end{align}
so the memory is maximized for systems that, from the binary's
viewpoint, are viewed edge-on ($\theta=\pi/2$) and is minimized for
systems that are viewed face-on ($\theta=0$). This fact is clearly
highlighted in Fig.~\ref{fig:MemoryObservers}, which shows what the whole
gravitational wave strain looks like for a binary black hole system
when evaluated at the point on the sky with
$(\theta,\phi)=(\pi/2,0)$. As can be seen, even for an ideal
orientation (as well as an ideal binary, i.e., equal mass and large
black hole spins in the direction of the orbital angular momentum),
the net memory that is induced on the gravitational wave detector is $\lesssim50\%$ of the magnitude of the full signal.

Furthermore, a unique challenge for observing memory in a real-world
detector is the fact that detectors do not analyze the GW strain in
the time domain, but rather the frequency domain. This is an issue
because the part of the strain that sources the memory, even though it
can have a large amplitude, is a low-frequency effect since it
resembles a step function and is not as oscillatory as the other modes
of the strain. We highlight this in Fig.~\ref{fig:MemoryFreq}, which shows
the Amplitude Spectral Density (ASD), i.e., the root of the Power
Spectral Density (PSD), of the gravitational wave strain evaluated at
$(\theta,\phi)=(\pi/2,0)$. We also show the contributions to the
strain from the $(2,\pm 2)$ modes and the energy flux. The NR system
is the same as that in Fig.~\ref{fig:MemoryObservers}. Before Fourier
transforming the waveform we first pre-process it following Ref.~\cite{Chen:2024ieh}, i.e., we subtract a linear function from the waveform to remove the offset due to the memory effect and we apply a
Planck window with an $\epsilon$ of $10^{-4}$~\cite{McKechan:2010kp}. As is shown, the ASD
of the strain is primarily represented by the $(2,\pm2)$ modes, while
the energy flux, i.e., the memory, only contributes to the ASD at
frequencies below 10Hz. This is what makes memory challenging to
observe in current ground-based detectors. Because of seismic noise,
LVK detectors are not sensitive to signals below $\sim$10Hz. In
particular, the signal-to-noise ratio (SNR) $\rho$ of the strain and
the memory contribution to the strain, i.e.,
\begin{align}
\rho=\sqrt{4\int_{f_{\mathrm{min}}}^{f_{\mathrm{max}}}\frac{|\tilde{h}(f)|^{2}}{S_{n}(f)}df},
\end{align}
where $\tilde{h}(f)$ is the Fourier transform of the strain,
$S_{n}(f)$ is the noise PSD, are $\approx$ 65 and $\approx$ 2. Because of this low SNR, current efforts to observe memory rely on a procedure called
``stacking'' which combine the SNR estimates of the memory from
various events to instead compute a type of population
measurement~\cite{Hubner:2019sly,Hubner:2021amk,Grant:2022bla}. Put
differently, these stacking efforts aim to show that memory has been
detected in a population of events. To detect memory in a single
event, we will most-certainly need to rely on future ground-based detectors
like the Einstein Telescope or Cosmic Explorer (CE) or perhaps even one of the planned
space-based detectors like LISA, which are more susceptible to the
lower frequency regimes.

\subsection{Memory Correction}
\label{sec:memorycorrection}

\begin{figure}[t]
  \includegraphics[width=0.5\textwidth]{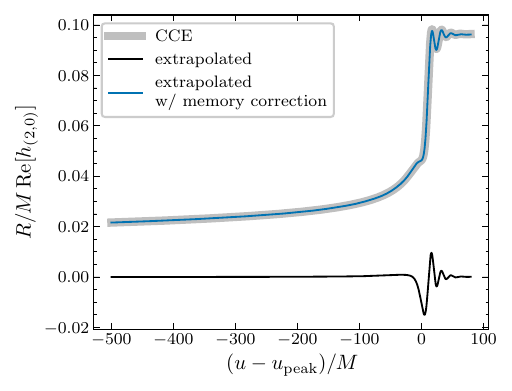}
  \caption{%
    Comparison of the real part of the $(2,0)$ mode of the strain
    extracted using CCE (gray) to that of the strain extracted using
    the extrapolation procedure (black) as well as the its
    memory-corrected version (blue). We apply the memory correction by
    computing the energy flux, Eq.~\eqref{eq:energyflux}, using the
    extrapolated strain waveform and then adding it to this
    waveform. The system used is the same as in Fig.~\ref{fig:ChargeFlux}.}
  \label{fig:EXTMemoryCorrection}
\end{figure}

Although we have so far primarily presented memory as being easily
resolvable in NR simulations, this is not the case. In particular,
prior to the creation of a CCE code framework by
Ref.~\cite{Pollney:2010hs} in 2010 and a more efficient framework by
Ref.~\cite{Mitman:2020pbt} in 2020, there was no memory in the
waveforms that were produced by NR simulations. This was because, when
not using CCE, NR simulations relied on extrapolation to produce
predictions for what the gravitational wave strain should be at null
infinity. However, because extrapolation is not an exact solution to
Einstein's equations, it is unable to accurately resolve memory
effects. This important fact is shown in Fig.~\ref{fig:EXTMemoryCorrection}, which
shows the $(2,0)$ mode of the strain for waveforms extracted using CCE
and extrapolation. As can be seen, the CCE waveform captures both the
memory and the oscillations induced in the merger and ringdown phases,
but the extrapolated waveform only captures the later.

Even so, Ref.~\cite{Mitman:2020bjf} showed that extrapolated waveforms
can be corrected, through a post-processing technique, to include the
memory contribution and exhibit much better agreement with
CCE waveforms.\footnote{Ref.~\cite{Talbot:2018sgr} also performed a correction to extrapolated waveforms to include memory, but did not compare the corrected waveforms to CCE waveforms, which naturally exhibit this effect.} In Ref.~\cite{Mitman:2020bjf}, it was found that for a
wide range of binary simulations, the extrapolated strain waveforms
simply seem to fail to capture the energy flux contribution in
Eq.~\eqref{eq:supertranslationconservationlaw}. Thus, the authors argued that
since the energy flux is only a function of the strain, the
extrapolated strain waveforms can be self-consistently corrected to
include the missing memory contribution. In doing so, they found that
with such a correction the extrapolated waveforms then satisfy the
supertranslation conservation law to a higher degree and better match
the CCE waveforms. For completeness, we also show this phenomenon in
Fig.~\ref{fig:EXTMemoryCorrection}. By eye, one can easily see that once the
extrapolated strain waveform is corrected to include the missing
energy flux contribution, it agrees much better with the CCE
waveform.

Unfortunately, even with this useful memory correction, Ref.~\cite{Mitman:2020bjf} found that extrapolated waveforms still do not outperform CCE waveforms in terms of their accuracy and violation of the Bianchi identities. Consequently, future waveform models should use CCE waveforms.

\section{BMS Frame Fixing}
\label{sec:BMSresults}

Up until this point, we have primarily been focused on utilizing the
charge/flux perspective enlightened through the BMS group to
understand memory effects and the means by which the gravitational wave strain is sourced. For the rest of this review, we will turn our
attention to how the transformations of the BMS group are crucial for
performing robust analyses with NR waveforms and building waveform
models to test Einstein's theory of GR with gravitational wave
detector data.

\subsection{Fixing the frame with BMS charges}

\label{sec:BMSintro}

As is the case in every field of physics, fixing the frame of the
system that one is studying is vitally important to ensure that the
observed phenomena are physical and not just gauge artifacts. For
binary black hole mergers, the situation is no different. In
particular, whenever we study the gravitational radiation from a
binary merger, we are often implicitly fixing part of the frame
without knowing it. For example, by specifying a $\hat{z}$-axis, e.g.,
the direction of the binary's orbital angular momentum or the
direction of the remnant black hole's spin axis, with respect to which
we construct spherical harmonics to decompose the gravitational wave
strain in, we are inherently fixing the rotation freedom by
constructing a canonical direction. However, while this aspect of the
frame fixing may seem trivial, there are other freedoms that need to
be fixed that are more subtle.

\begin{figure}[tb]
  \includegraphics[width=0.5\textwidth]{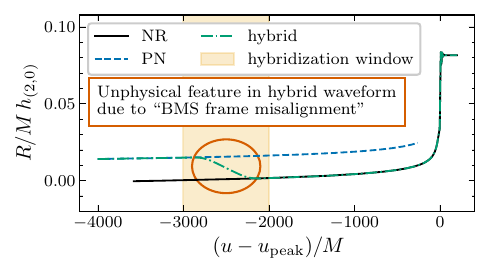}
  \caption{%
    The $(2,0)$ mode of the strain waveform computed in a NR system
    (solid/black), a PN system (dashed/blue), and the corresponding
    NR/PN hybrid waveform (dash-dotted/green). The hybridization
    window is the highlighted region in yellow. An unphysical feature
    in the NR/PN hybrid waveform as a result of an improper BMS frame
    alignment is circled in red. The system is the same as that in
    Figs.~\ref{fig:ChargeFlux} and~\ref{fig:EXTMemoryCorrection}.}
  \label{fig:BMSFrameIssue}
\end{figure}

When fixing the frame of a system, one needs to fix the
transformations that are contained within the system's symmetry
group. For gravitational radiation, in which the symmetry group is the
BMS group, this means that to fix the frame of, say, the waveform
radiated to future null infinity in a binary black hole merger, one
must not only fix the rotation freedom, but also the boost and the
supertranslation freedom. But, for an arbitrary system, how should
these freedoms be fixed? In principle, there is no canonical frame,
since GR has no preferred frame. In practice, however, whatever
analysis one would like to conduct on a system often has its own
canonical frame, in the sense that there is a certain frame which
makes the analysis much simpler. For example, when using the
quadrupole formula to compute the gravitational wave emitted by a
binary system, it tends to be easier to do the calculation in the
frame in which the binary system's orbital angular momentum is aligned
with the $\hat{z}$-axis. If a different orientation is used, then
certain simplifications that would otherwise occur do not and one is
instead met with many more terms than are necessary to explain the
underlying physics. For studying numerical simulations, the same holds
true. In particular, because working with NR results often consists of
comparing NR waveforms to certain perturbative solutions to Einstein's
equations, like post-Newtonian (PN) in the early inspiral phase or
black hole perturbation theory (BHPT) in the ringdown, the ideal frame
is typically set by the perturbative result. Then, to map the NR
system to this canonical frame one must work with the charges
corresponding to the frame freedoms, i.e., the BMS charges in
Eqs.~\eqref{eq:Poincarecharges} and
Eq.~\eqref{eq:supermomentumcharge}, and find the transformation which
maps these charges in the NR system to the values expected by the
canonical frame. For example, to map to the system's center-of-mass
frame, one can simply take the translation and boost charges in
Eqs.~\eqref{eq:Poincarecharges} and find the translation and boost
which maps these charges to zero.

While at first this business of fixing the frame may seem trivial, we
stress that performing this procedure is of the utmost importance to
extract meaningful physics. To help highlight this, we present
Fig.~\ref{fig:BMSFrameIssue}, which shows the consequence of not accounting for
the difference in frames between a NR and a PN system. One important
aspect of waveform modeling is constructing waveforms which span the
entire frequency band of GW detectors. This means that one must build
waveforms that contain thousands of orbits. NR simulations, however,
typically only contain tens or perhaps hundreds of orbits, but not
nearly as many as required to span the frequency bands of current and
future observatories. Therefore, there is a constant need to
``hybridize'' finite NR waveforms with perturbative solutions that are
accurate during the early inspiral regimes and can be more efficiently
calculated. Often these hybrid waveforms are built by hybridizing
finite NR waveforms with PN waveforms. This results in waveforms which
are PN during the early inspiral phase, a smooth blend of PN and NR
over a window called the hybridization window, and NR for the
remainder of the binary coalescence. As is shown by
Fig.~\ref{fig:BMSFrameIssue}, however, if one does not account for the freedoms
resulting from the BMS group, i.e., if one does not map the NR system
to be in the same BMS frame as the PN system, then this hybridization
procedure will fail in the sense that it will introduce unphysical
features in the hybrid waveform. Fortunately, this issue can be
mitigated rather easily by simply mapping the NR system to be in the
right frame. But, in order to do so, one needs to know how the various
BMS symmetries transform the asymptotic coordinates as well as the
asymptotic data of interest.

Fortunately, this was already presented in Sec.~\ref{sec:BMSeffects}.
With both Eq.~\eqref{eq:coordinatetransformation} and
Eqs.~\eqref{eq:datatransformation}, one has all of the information
that is needed to transform asymptotic data. The only ingredient that
remains for fixing the frame of the asymptotic data is which charges
one should use to constrain the BMS freedoms.\footnote{Note that throughout this review we ignore the subtle fact that the BMS charges presented in Eqs.~\eqref{eq:Poincarecharges} and in the remainder of this section are not supertranslation-invariant~\cite{Christodoulou:1991cr,Chen:2021kug,Javadinezhad:2022hhl,Javadinezhad:2022ldc,Javadinezhad:2023mtp}.} Normally, one could
just use the charges presented in Eqs.~\eqref{eq:Poincarecharges}
and Eq.~\eqref{eq:supermomentumcharge}. However, because it is often
the case that one wishes to compare NR systems to PN predictions for
which there is only the strain and not the Weyl scalars, it turns out
that there are more convenient BMS charges that can be used. In
particular, rather than working with the translation and boost
charges, i.e., Eq.~\eqref{eq:momentumcharge} and
Eq.~\eqref{eq:boostcharge}, it is useful to construct the
center-of-mass charge~\cite{Kozameh:2013bha,Compere:2019gft}
\begin{align}
\label{eq:comcharge}
G^{a}(u)&=(K^{a}+uP^{a})/P^{t}\nonumber\\
&=\frac{1}{4\pi}\int_{S^{2}}\mathrm{Re}\left[\left(\bar{\eth}n^{a}\right)\left(\hat{N}+u\eth m\right)\right]\,d\Omega/P^{t}.
\end{align}
This measures the spacetime's center-of-mass motion.\footnote{One can think of the boost charge $K^{a}$ as measuring the translation away from the origin and the $uP^{a}$ term as measuring the change in center-of-mass due to a non-zero velocity.} It is useful since we will always want to examine
systems in their center-of-mass frame, which can be mapped to by
finding the translation and boost which minimize
Eq.~\eqref{eq:comcharge}. As for the rotation charge, again because we
typically do not have access to either of the $\Psi_{1}$ or
$\Psi_{2}$ Weyl scalars in PN theory, one typically utilizes a
rotation-like charge that can be computed from the strain. In
particular, in Ref.~\cite{Boyle:2013nka} such a rotation charge was built by finding
the angular velocity which keeps the radiative fields, e.g., the
strain, as constant as possible in the corotating frame of the
binary system. This ``charge'' is
\begin{align}
\label{eq:angularvelocity}
\vec{\omega}(u)=-\langle\vec{L}\vec{L}\rangle^{-1}\cdot\langle\vec{L}\partial_{t}\rangle,
\end{align}
where
\begin{subequations}
\begin{align}
\langle\vec{L}\partial_{t}\rangle^{a}&\equiv\sum\limits_{\ell,m,m'}\mathrm{Im}\left[\bar{f}_{(\ell,m')}\langle\ell,m'|L^{a}|\ell,m\rangle\dot{f}_{(\ell,m)}\right],\\
\langle\vec{L}\vec{L}\rangle^{ab}&\equiv\sum\limits_{\ell,m,m'}\bar{f}_{(\ell,m')}\langle\ell,m'|L^{(a}L^{b)}|\ell,m\rangle f_{(\ell,m)},
\end{align}
\end{subequations}
and $f(u,\theta,\phi)$ is some function corresponding to the
asymptotic radiation, such as the GW strain or the news. In
Eq.~\eqref{eq:angularvelocity}, $\vec{L}$ is the infinitesimal
generator of rotations.

When the system only consists of one black hole, as is the case, for
example, when studying the ringdown phase, this prescription can break
down. In this case, it is more useful to fix the rotation freedom
using the intrinsic spin of the black hole, to ensure that one is in
the frame of the individual black hole. To compute this, one can
either map to the center-of-mass frame of the black hole and compute
the angular momentum charge, or, for simplicity, they can instead
compute~\cite{Iozzo:2021vnq}
\begin{align}
\label{eq:spincharge}
\vec{\chi}(u)=\frac{\gamma}{M_{B}^{2}}\left(\vec{J}+\vec{v}\times\vec{K}\right)-\frac{\gamma-1}{M_{B}^2}\left(\hat{v}\cdot\vec{J}\right)\hat{v},
\end{align}
which achieves the same result. Here
\begin{align}
\gamma(u)\equiv\left(1-|\vec{v}|^{2}\right)^{-1/2}
\end{align}
is the Lorentz factor,
\begin{align}
M_{B}(u)\equiv\sqrt{-\eta_{\mu\nu}P^{\mu}P^{\nu}}
\end{align}
is the Bondi mass,
\begin{align}
\vec{v}(u)\equiv\vec{P}/P^{t}
\end{align}
is the velocity vector, and the vectors $\vec{J}$ and $\vec{K}$ are
the angular momentum and boost charges found in
Eq.~\eqref{eq:angularmomentumcharge} and
Eq.~\eqref{eq:boostcharge}. With this ``charge'', one can then map to
the frame of the black hole by, say, finding the rotation which aligns
this charge with the positive $\hat{z}$-axis. Again, we stress that Eq.~\eqref{eq:angularvelocity} and Eq.~\eqref{eq:spincharge} have been introduced to help with fixing the rotation freedom by constructing a preferred axis, either through the angular velocity vector of a PN waveform or the spin of an isolated black hole.

For fixing the supertranslation, since for fixing the Poincaré
transformations we have been able to rely on the charges in
Eqs.~\eqref{eq:Poincarecharges}, one may naively think that we can simply use the supermomentum charge in
Eq.~\eqref{eq:supermomentumcharge}. This charge, however, turns out to be
supertranslation-invariant in nonradiative regimes of future null
infinity, i.e., regimes where the news is zero. As a result, it
cannot be used to fix the supertranslation freedom.\footnote{
	This phenomenon should not necessarily come as a surprise, as this
	also happens for the translation-invariant momentum charge.}
Instead, one must construct a different supertranslation ``charge'' which transforms in
a meaningful way. In Ref.~\cite{OMMoreschi_1986,Moreschi:1988pc,Moreschi:1998mw,Dain:2000lij}, such a charge was presented. It is
simply Eq.~\eqref{eq:generalsupermomentum} with $p=1$ and $q=0$ and is the Moreschi supermomentum
\begin{align}
\label{eq:Moreschisupermomentum}
\mathcal{P}_{\mathrm{M}}(u,\theta,\phi)&\equiv\Psi_{2}+\sigma\dot{\bar{\sigma}}+\eth^{2}\bar{\sigma}\nonumber\\
&=-m+\mathrm{Re}\left[\eth^{2}\bar{\sigma}\right]\nonumber\\
&=\int_{-\infty}^{u}|\dot{\sigma}|^{2}\,du-M_{\mathrm{ADM}},
\end{align}
where $M_{\mathrm{ADM}}$ is the ADM mass~\cite{Arnowitt:1959ah}. An important property of the Moreschi supermomentum is that, because
it transforms as
\begin{align}
\label{eq:Moreschisupermomentumtransformation}
\mathcal{P}_{\mathrm{M}}'&=\frac{1}{k^{3}}(\mathcal{P}_{\mathrm{M}}-\eth^{2}\bar{\eth}^{2}\alpha)\nonumber\\
&=\frac{1}{k^{3}}\left(-m+\mathrm{Re}\left[\bar{\eth}^{2}\left(\sigma-\eth^{2}\alpha\right)\right]\right),
\end{align}
in Minkowski space where $m=0$, minimizing the Moreschi supermomentum
corresponds to finding the supertranslation that ensures the spacetime
is really Minkowski, not supertranslated Minkowski. Furthermore,
this property is true so long as the $\ell\geq2$ components of
the mass aspect are zero, which is the same as there being no Geroch
supermomentum. This is useful because isolated black holes cannot
have Geroch supermomentum, so if one instead uses the Moreschi
supermomentum to fix the BMS freedom, they ensure that they are always
working in the frame that is the most natural frame from the
perspective of the individual black hole. We will call the BMS frame in which
the Moreschi supermomentum has zero-valued $\ell\geq1$ modes the \emph{superrest frame}, as it is in some sense
an extension of the notion of a rest frame. To map to such a frame,
one can simply solve
Eq.~\eqref{eq:Moreschisupermomentumtransformation} for
$\mathcal{P}_{M}'=-M_{B}$, which yields
\begin{align}
\label{eq:nicesection}
\eth^{2}\bar{\eth}^{2}\alpha=\mathcal{P}_{M}(u=\alpha,\theta,\phi)+k_{\mathrm{rest}}(\alpha,\theta,\phi)^{3}M_{B}(\alpha)
\end{align}
where $k_{\mathrm{rest}}$ is a special case of the conformal factor in
Eq.~\eqref{eq:conformalfactor} in the sense that it is the conformal
factor for a boost whose velocity matches the momentum charge:
\begin{align}
k_{\mathrm{rest}}(u,\theta,\phi)\equiv\frac{1}{\gamma\left(1-\vec{v}\cdot\vec{r}\right)}=\frac{M_{B}}{P_{a}n^{a}}.
\end{align}
Furthermore, in Ref.~\cite{Dain:2000lij}, the authors proved that, for a
condition on the energy flux, which is always obeyed in nonradiative
regimes of future null infinity, Eq.~\eqref{eq:nicesection} always has
a regular solution. It was also shown that Eq.~\eqref{eq:nicesection}
can be solved iteratively. That is, if one wishes to find the
supertranslation that maps a system to the superrest frame at some
time $u_{0}$, they can evaluate the right-hand side of
Eq.~\eqref{eq:nicesection} at time $u=u_{0}$, solve for $\alpha$,
evaluate the right-hand side at time $u=\alpha$, solve for a new
$\alpha$, etc., until $\alpha$ converges to a solution. This is the
underpinning of the frame fixing program. Specifically, by utilizing
this fact, one can always solve for the infinite degrees of freedom in
the BMS transformation and map to a target BMS frame in an iterative
fashion. For example, to map a NR system to the PN BMS frame, one can
\begin{enumerate}
  \item Construct a window during the early inspiral phase over which
    to align a NR system to a PN system;
  \item Iteratively solve Eq.~\eqref{eq:nicesection} for the
    supertranslation which maps the $\ell\geq2$ components of the
    difference of the NR/PN Moreschi supermomenta to zero; this is
    equivalent to mapping to the superrest frame at
    $u\rightarrow-\infty$, i.e., in the infinite past of the binary;
  \item Find the frame rotation that maps the NR system's angular
    velocity vector (see Eq.~\eqref{eq:angularvelocity}) to match that
    of the PN system;
  \item Find the translation and boost which minimize the
    center-of-mass charge (see Eq.~\eqref{eq:comcharge});
  \item Perform a time and phase translation optimization to align the
    NR and PN waveforms in the window;
  \item Repeat steps (i) - (v), until the error between the NR and PN
    waveforms converges.
\end{enumerate}
For mapping to the frame of an individual black hole, like a remnant
black hole, the process is identical, but in step (ii) one can
minimize the Moreschi supermomentum, since the target Moreschi
supermomentum is $-M_{B}$ and in step (iii) the rotation charge should
instead be the spin of the black hole (see
Eq.~\eqref{eq:spincharge}). Also, step (v) is no longer
necessary since there is no preferred way to fix the time or phase freedom from the perspective of the remnant black hole. In the subsequent sections,
we explain in more detail how to map a system to either the PN
BMS frame or the superrest frame of an individual black hole and
provide some numerical results showing why frame fixing is important
and the success of this procedure at mapping NR systems to some reasonable BMS frame.

\subsection{PN BMS frame}

At the end of Sec.~\ref{sec:BMSintro} the steps for mapping to the PN
BMS frame through an iterative process using the BMS charges was
outlined. Ultimately, to map to this frame one must simply find the
BMS transformation which maps the charges of the NR system to match
those of the PN system. This, however, requires that one has knowledge
of the charges in PN theory. Unfortunately, most PN calculations focus
on computing the strain and not the Weyl scalars, so the majority of
the canonical BMS charges can not be computed in a PN
formulation. Therefore, one can either compute these charges in PN or
try to use alternative charges that are functions of only the
strain. Because the later is easier, this is what is typically used in
BMS frame fixing. But, even if one uses charges that are only
functions of the strain, these charges must still be computed in
PN. Fortunately, the center-of-mass charge, which is zero for PN
systems, and the angular velocity vector, which can easily be computed
numerically from the PN strain, are painless to obtain. The PN
Moreschi supermomentum, however, because it involves a time integral,
cannot be computed numerically and must instead be worked out
analytically. This was first performed in Ref.~\cite{Mitman:2022kwt},
which computed it from the PN strain using
Eq.~\eqref{eq:Moreschisupermomentum} to 3PN order
without spins and 2PN order with spins. They then implemented this
iterative procedure for fixing the frame of NR systems in the publicly
available python module \texttt{scri}~\cite{scri}.

\begin{figure}[ht!]
  \includegraphics[width=0.5\textwidth]{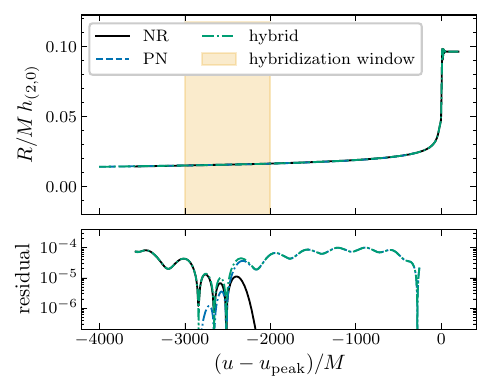}
  \caption{%
    The $(2,0)$ mode of the strain waveform computed in a NR
    system (solid/black), a PN system (dashed/blue), and the
    corresponding NR/PN hybrid waveform (dash-dotted/green). The
    hybridization window is the highlighted region in yellow. In the
    bottom panel, we show he absolute error between the NR and PN
    waveforms (green), the hybrid and NR (black), and the hybrid and
    the PN (blue). The NR simulation is the same as that used in
    Figs.~\ref{fig:ChargeFlux},~\ref{fig:EXTMemoryCorrection}, and~\ref{fig:BMSFrameIssue}.%
  }
  \label{fig:BMSFrameIssueResolved}
\end{figure}

In Fig.~\ref{fig:BMSFrameIssueResolved}, we demonstrate the overall success of the BMS
frame fixing procedure by mapping a NR system to the PN BMS frame and
computing the error between the NR waveform and the PN waveform in
this frame. This figure is identical to Fig.~\ref{fig:BMSFrameIssue}, but
correctly utilizes the BMS freedoms at future null infinity to perform
the waveform alignment. As can be seen, by mapping the NR waveform to
the PN BMS frame, the absolute error between the two waveforms is
decreased by three orders of magnitude. Also, as seen through the
hybrid waveform built from these two inputs, by fixing the frame
properly the hybrid waveform no longer has an unphysical feature that
could bias data analyses that used this waveform. Furthermore, we
stress that although here the effects of fixing the frame are a bit
pronounced due to us studying the $(2,0)$ mode, if one instead studies
other modes they will still find that the absolute error is improved.

\subsection{Superrest frame}

Apart from comparing NR systems to PN predictions to construct more
effective hybrid waveforms and models, NR simulations of black holes
are also particularly useful for extracting the quasi-normal mode
(QNM) amplitudes expected by perturbation theory~\cite{Teukolsky:1973ha,Berti:2009kk}. A subtle
issue, however, is that when a perturbed black hole is formed, either
through a black hole merger or stellar collapse, it is not in the
frame that black hole perturbation theory is typically performed
in. Put differently, the remnant black hole formed in some
astrophysical event may not be described by the usual Kerr metric, but
instead a boosted or a supertranslated version. Consequently, if one
tries to study the perturbations of the remnant black hole formed in
a NR simulation without accounting for the difference in BMS frames,
then the analysis will fail in the same way that the hybridization
shown in Fig.~\ref{fig:BMSFrameIssue} fails.

\begin{figure}[ht!]
  \includegraphics[width=0.5\textwidth]{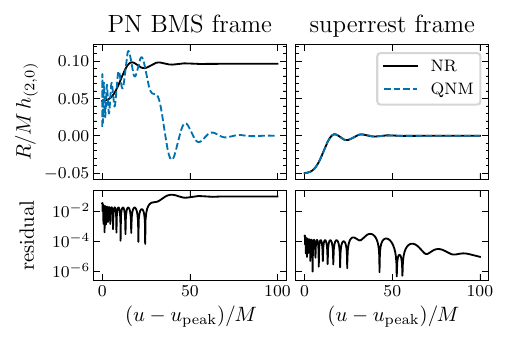}
  \caption{%
    \emph{Top row:} The $(2,0)$ mode of the strain waveform for a NR
    system (black) and the best fit QNM model (blue). The QNM model
    contains seven $(2,0)$ overtones. On the left, the NR waveform is
    in the PN BMS frame, while on the right, it is instead in the
    superrest frame of the remnant black hole. %
    \emph{Bottom row:} Residual between the NR waveform and the fit.}
  \label{fig:RemnantFrame}
\end{figure}

In particular, in Fig.~\ref{fig:RemnantFrame}, we show exactly this. In the
top row we show two fits of QNMs to the $(2,0)$ mode of the NR strain
waveform from the simulation shown in
Figs.~\ref{fig:ChargeFlux},~\ref{fig:EXTMemoryCorrection}, and~\ref{fig:BMSFrameIssue}. Meanwhile,
in the bottom row we show the residuals. The QNM model that we
consider has seven overtones and both mirror modes.\footnote{%
  We stress that here we use this model only for simplicity and to
  show the importance of fixing the frame. In reality, when trying to
  correctly fit a NR waveform with QNMs, one should take precaution
  to ensure that the QNMs are meaningfully resolved.} %
As seen through the left-most column, which corresponds to the NR
waveform in the PN BMS frame, i.e., the wrong frame for BHPT analyses,
when the NR system is not in the superrest frame of the remnant black
hole the strain can not be fit with QNMs, in part because of the
nonzero asymptotic value resulting from the memory effect. But, if one
maps the NR system to the superrest frame of the remnant black hole,
then the QNM fit can be performed successfully and the residual is
reduced by three orders of magnitude. This illustrates that fixing the
frame is essential for extracting physical QNM amplitudes.

Another important result that can be seen through Fig.~\ref{fig:RemnantFrame}
and the bottom-left panel of Fig.~\ref{fig:ChargeFlux} is an example of the
nonlinear nature of GR and a highlight of the need for predictions
from second-order perturbation theory to study the ringdown of NR
waveforms. More specifically, since a large portion of the $(2,0)$
mode of the strain is sourced by the energy flux, i.e.,
Eq.~\eqref{eq:energyflux}, this means that first-order perturbation
theory is not sufficient to explain the ringdown excitations of the
strain near the point of peak luminosity $u_{\mathrm{peak}}$. In
particular, because the energy flux is inherently nonlinear, i.e., it
goes as the strain squared, this means it can only be modeled by
second-order black hole perturbation theory, e.g., by the quadratic
QNMs studied in Refs.~\cite{Mitman:2022qdl,Cheung:2022rbm}.

\section{Discussion}
\label{sec:discussion}

In this review, we have presented a vast amount of information regarding memory effects and the BMS group. In particular, we began in Sec.~\ref{sec:BMSmotivation} by studying where in a NR simulation one should extract gravitational wave information for gravitational wave detectors. While it may be tempting to extract GWs on some timelike wordline, like that of an inertial detector, we showed how this can quickly lead to coordinate ambiguities and thus motivated the need to extract GWs at future null infinity $\mathcal{I}^{+}$---the final destination of outgoing radiation. By working with data on $\mathcal{I}^{+}$, instead of in the bulk, one is free from the full diffeomorphism invariance of GR and obtains a more suitable set of coordinates to work with. However, as we outlined in Sec.~\ref{sec:Bondigauge}, even on $\mathcal{I}^{+}$ that are still troubling coordinate ambiguities that are present because of the symmetry group of null infinity: the BMS group. Consequently, to work with data on $\mathcal{I}^{+}$, one must have a robust understanding of the BMS group and the way in which it transforms coordinates and asymptotic data. 

In Sec.~\ref{sec:BMSgroup} we reviewed the infamous BMS group; specifically how it can be viewed as the Poincaré group, but with the usual spacetime translations replaced by an infinite group of transformations called \emph{supertranslations}. These can be viewed as direction-dependent translations and arise from the fact that each point on $\mathcal{I}^{+}$, which can each be thought of as a single observer, ends up being casually disconnected from any other. Then, in Sec.~\ref{sec:BMSeffects}, we showed how an arbitrary BMS transformation changes the spacetime coordinates that the asymptotic data will be functions of. We also explained what this data on $\mathcal{I}^{+}$ should be, i.e., not only the gravitational wave strain, but also the five complex Weyl scalars, which encode information about the spacetime curvature. 

Finally, in Sec.~\ref{sec:memoryoverview}, we used our understanding of the symmetries of $\mathcal{I}^{+}$ to construct an intuition for why memory effects should exist in GR. In particular, because of Neother's theorem~\cite{Wald:1999wa}, one might expect that the supertranslation symmetries of $\mathcal{I}^{+}$ yield some kind of conservation law between an angle-dependent mass and an angle-dependent energy flux, like that of Eq.~\eqref{eq:nearlypedagogicalconservationlaw}. However, because changes to the $\ell\geq2$ harmonics of system's mass multipole moment source GWs, one needs an extra ingredient on the left-hand side of Eq.~\eqref{eq:nearlypedagogicalconservationlaw}, i.e., that of Eq.~\eqref{eq:pedagogicalconservationlaw}, to balance this fact. Therefore, the conservation law that stems from supertranslations, Eq.~\eqref{eq:pedagogicalconservationlaw} or Eqs.~\eqref{eq:conservationlaws}, is really a statement about the memory that is induced on the spacetime by any system that emits gravitational radiation.

After our pedagogical overview, in Sec.~\ref{sec:literaturereview} we reviewed the literature behind memory effects and the BMS group. Focusing on the memory first, we highlighted how this effect has been known since 1974, but did not receive much attention until its connection to asymptotic symmetries and soft theorems was found in 2013 and 2014~\cite{Strominger:2013jfa,Winicour:2014ska,Flanagan:2014kfa,Strominger:2014pwa}. Ever since this realization, there has been a surge of interest and discovery in tangentially related fields, such as celestial holography~\cite{Strominger:2017zoo, Pasterski:2019ceq,Raclariu:2021zjz, Pasterski:2021rjz, Pasterski:2021raf}.

In Sec.~\ref{sec:mathoverview}, we presented the more formal mathematics behind our earlier discussions. In particular, we showed how the Bondi-Sachs metric and Einstein's equations yield a series of evolution equations for both the mass and angular momentum aspects (see Eqs.~\eqref{eq:evolution}), which, when integrated with respect to time, yield the formal version of the pedagogical conservation law of Eq.~\eqref{eq:pedagogicalconservationlaw}, i.e., Eqs.~\eqref{eq:conservationlaws}. These equations show that the real and imaginary parts of the gravitational wave strain (or shear) can be written in terms of mass and angular momentum charges and fluxes, which help illustrate the different ways in which the strain or memory can be sourced. That is, Eqs.~\eqref{eq:conservationlaws} present the modern interpretation of memory: it can be sourced by a change in the system's charges (\emph{linear} or \emph{ordinary} memory) or a change in the fluxes (\emph{nonlinear}, \emph{Christodoulou}, or \emph{null} memory).

In Sec.~\ref{sec:memoryresults}, we implemented our mathematical results numerically for simulations of binary black hole mergers. In Sec.~\ref{sec:conservationlawresults} we took Eqs.~\eqref{eq:conservationlaws} and evaluated them for a GW150914-like NR simulation in Fig.~\ref{fig:ChargeFlux}. This showed how useful Eqs.~\eqref{eq:conservationlaws} can be for studying NR simulations and extracting the memory contribution to the strain, as highlighted in the lower left panel of Fig.~\ref{fig:ChargeFlux}. In Sec.~\ref{sec:detectability}, in Fig.~\ref{fig:MemoryFreq} we showed what the frequency spectrum of a typical memory signal looks like relative to the LIGO and CE amplitude spectral densities. As shown, even though memory is fairly pronounced in the time domain (see, e.g., Fig.~\ref{fig:MemoryObservers}), in the frequency domain memory occupies the lower part of the band which makes it challenging to detect in single events with current detectors. However, by stacking events and trying to detect memory in a population~\cite{Hubner:2019sly,Hubner:2021amk,Grant:2022bla} or by relying on future detectors like CE, ET, or LISA~\cite{Islo:2019qht,Grant:2022bla}, the chances for observing memory for the first time are much higher.

Last, in Sec.~\ref{sec:BMSresults} we highlighted the importance of controlling the BMS freedoms for studying NR waveforms. First, in Fig.~\ref{fig:BMSFrameIssue}, we showed how NR and PN waveforms naturally disagree with one another, due to the fact that PN waveforms have information about the past history of the compact binary's inspiral, while numerical simulations do not. Furthermore, Fig.~\ref{fig:BMSFrameIssue} shows that, without resolving this issue, if one tries to construct a hybrid waveform which is needed for building waveform models for the LVK collaboration, then the hybrid waveform exhibits an unphysical feature due to this disagreement. Fortunately, this disagreement simply turns out to be an issue of the two waveforms being in different BMS frames, i.e., they differ by some arbitrary BMS transformation. In the text surrounding Fig.~\ref{fig:BMSFrameIssue}, we outlined how one can use BMS charges to determine this transformation and fix the BMS frame of the NR waveform to match that of the PN waveform. After doing so, one obtains Fig.~\ref{fig:BMSFrameIssueResolved} which shows much better agreement between NR and PN strain waveforms. Finally, in Fig.~\ref{fig:RemnantFrame}, we also highlighted why fixing the BMS frame is important for studying the ringdown phase of NR simulations with black hole perturbation theory. In particular, because of the memory, the remnant black hole formed in a black hole merger is not exactly Kerr, but \emph{supertranslated Kerr}. Therefore, if one wishes to fit the ringdown phase with QNMs, they need to find the BMS transformation that maps the remnant black hole to the same frame of the usual Kerr metric: the \emph{superrest} frame. After doing so, one can then perform meaningful ringdown fits and analyses, as is illustrated through the right-hand panels of Fig.~\ref{fig:RemnantFrame}.

 It is our hope that this review will serve as a resource for those interested in learning more about memory effects and the BMS group in the context of NR simulations and GW modeling more generally. As new and more accurate GW detectors come online, understanding the complexities of the GWs that GR predicts, such as those we have covered here, will be of the utmost importance, as they will likely be key to observing never-before-seen physics that will futher illuminate how our Universe works.

\section*{Acknowledgments}
\label{sec:acknowledgements}

The authors thank Eanna Flanagan for informative conversations about
the history of memory effects and Alexander Grant for intuition behind
BMS balance laws. Computations for this work were performed with the
Wheeler cluster at Caltech.
This work was supported by the Sherman Fairchild Foundation and NSF
Grants No. PHY-2011968, PHY-2011961, PHY-2309211, PHY-2309231,
OAC-2209656 at Caltech.  The work of L.C.S. was partially supported by
NSF CAREER Award PHY-2047382 and a Sloan Foundation Research
Fellowship.

\appendix

\section{Expressing a BMS transformation in terms of rotations, boosts, and supertranslations}
\label{sec:BMSdecomposition}

In this Appendix, we briefly outline how one can write an arbitrary BMS transformation as a supertranslation followed by a Lorentz transformation. This is useful for mapping NR waveforms to a particular BMS frame because it enables one to compose BMS transformations. We begin by studying Lorentz transformations.

\subsubsection{$\mathrm{SL}(2,\mathbb{C})$ representation of a Lorentz transformation}

Let us begin with rotations. For a rotation $R$ by an angle $\theta$ about the axis $\hat{r}=\left(r_{x},r_{y},r_{z}\right)$, one may write this as the quaternion\footnote{For a review of quaternions, see Ref.~\cite{Doran:2007tqa}.}
\begin{align}
\mathbf{q}&=\mathrm{exp}\left(\frac{1}{2}\theta\hat{r}\right)\nonumber\\
&=\cos(\theta/2)\mathbf{I}+\left(r_{x}\mathbf{i}+r_{y}\mathbf{j}+r_{z}\mathbf{k}\right)\sin(\theta/2),
\end{align}
where $\mathbf{I}$, $\mathbf{i}$, $\mathbf{j}$, $\mathbf{k}$ are the elementary quaternions obeying the usual multiplication rules. Using spin matrices, i.e., elements of $\mathrm{SL}(2,\mathbb{C})$, one may write these quaternions as
\begin{align}
\boldsymbol{\sigma}&\equiv\left(\mathbf{I},\mathbf{i},\mathbf{j},\mathbf{k}\right)\nonumber\\&=\left(\begin{pmatrix}1&0\\0&1\end{pmatrix},\begin{pmatrix}0&i\\i&0\end{pmatrix},\begin{pmatrix}0&-1\\1&0\end{pmatrix},\begin{pmatrix}i&0\\0&-i\end{pmatrix}\right).
\end{align}
Therefore, the rotation $R$ has the $\mathrm{SL}(2,\mathbb{C})$ representation
\begin{align}
\label{eq:rotationrep}
\tilde{R}=\left(\cos(\theta/2),\hat{r}\sin(\theta/2)\right)\cdot\boldsymbol{\sigma}.
\end{align}
Note that in Eq.~\eqref{eq:rotationrep} $\tilde{R}$ is a unitary matrix. Meanwhile, for boosts, since a boost $B$ by rapidity $w$ along the axis $\hat{v}=(v_{x},v_{y},v_{z})$ is nothing more than a rotation by the angle $iw$ about the axis $\hat{v}$, a boost has the representation
\begin{align}
\label{eq:boostrep}
\tilde{B}=\left(\cosh(w/2),i\hat{v}\sinh(w/2)\right)\cdot\boldsymbol{\sigma}.
\end{align}
Note that in Eq.~\eqref{eq:boostrep} $\tilde{B}$ is a Hermitian
matrix. Because $\tilde{L}$---the $\mathrm{SL}(2,\mathbb{C})$ representation of some Lorentz transformation---is invertible, it has a unique polar
decomposition.  Correspondingly, a general Lorentz transformation
can be written as a composition of a boost followed by a rotation,
which in the $\mathrm{SL}(2,\mathbb{C})$ representation is
\begin{align}
\tilde{L}&=\tilde{R}\cdot\tilde{B}\nonumber\\
&=\left[\left(\cos(\theta/2),\hat{r}\sin(\theta/2)\right)\cdot\boldsymbol{\sigma}\right]\nonumber\\
&\phantom{=.}\cdot\left[\left(\cosh(w/2),i\hat{v}\sinh(w/2)\right)\cdot\boldsymbol{\sigma}\right].
\end{align}
Note though that what is shown above is for a \emph{passive} Lorentz transformation. Put simply, the transformation is acting on the coordinates themselves. Under $L$, our original coordinates $X$ are transformed to an intermediate coordinate system $X'$ by a boost, and then to a final coordinate system $X''$ by a rotation. The parameters of the rotation should be interpreted as \emph{a rotation in the $X'$ coordinate system}, not in the $X$ coordinate system.

\subsubsection{Decomposing an $\mathrm{SL}(2,\mathbb{C})$ matrix
  into rotation and boost}

Because $\mathrm{SL}(2,\mathbb{C})$ is a double cover of the restricted Lorentz group $\mathrm{SO}^{+}(3,1)$, we can always write the $\mathrm{SL}(2,\mathbb{C})$ representation of a Lorentz transformation $L$ as
\begin{align}
\tilde{L}\equiv\begin{pmatrix}a&b\\c&d\end{pmatrix}
\end{align}
for some $a,b,c,d\in\mathbb{C}$ with $ad-bc=1$.
We would like to perform a polar decomposition as either
$\tilde{L}=\tilde{R}\cdot\tilde{B}$ or
$\tilde{L}=\tilde{B}'\cdot\tilde{R}'$, where $\tilde{R}$ or
$\tilde{R}'$ is unitary and thus a rotation, and where $\tilde{B}$ or
$\tilde{B}'$ is hermitian and thus a boost, as seen in
Eqs.~\eqref{eq:rotationrep} and~\eqref{eq:boostrep}.
This is easily accomplished
with a singular value decomposition,
\begin{align}
\tilde{L}=U\cdot\Sigma\cdot V^{\dagger},
\end{align}
where $U$ and $V$ are complex unitary matrices and $\Sigma$ is
diagonal, and in this case Hermitian. Using the inverse properties of $U$ and $V$, we can write $\tilde{L}$ in the more suggestive form
\begin{align}
\tilde{L}=\left(U\cdot V^{\dagger}\right)\cdot\left(V\cdot\Sigma\cdot V^{\dagger}\right)
\end{align}
or
\begin{align}
\tilde{L}=\left(U\cdot\Sigma\cdot U^{\dagger}\right)\cdot\left(U\cdot V^{\dagger}\right),
\end{align}
which give our desired decompositions,
\begin{align}
\tilde{L}=\tilde{R}\cdot\tilde{B}\,\,\text{with}\,\,\tilde{R}=U\cdot V^{\dagger}\,\,\text{and}\,\,\tilde{B}=V\cdot\Sigma\cdot V^{\dagger}
\end{align}
or
\begin{align}
\tilde{L}=\tilde{B}'\cdot\tilde{R}'\,\,\text{with}\,\,\tilde{B}'=U\cdot\Sigma\cdot U^{\dagger}\,\,\text{and}\,\,\tilde{R}'=U\cdot V^{\dagger}.
\end{align}

\subsubsection{Composition of BMS elements}

Equipped with the $\mathrm{SL}(2,\mathbb{C})$ representation of
arbitrary Lorentz transformations, we also want a decomposition of BMS
group elements as compositions of supertranslations followed by
Lorentz transformations, or vice versa.  This will make it
computationally convenient to compose BMS transformations.  Such a
decomposition is possible since supertranslations $\mathbb{T}$ are a normal
subgroup of the BMS group $\mathfrak{B}$.
Namely, we can construct the homomorphism to the quotient group
$\varphi:\mathfrak{B}\rightarrow\mathfrak{B}/\mathbb{T}\simeq\mathrm{SO}(3,1)$. We
can say that a BMS group element $s$ is a ``pure supertranslation'' if
$s\in \mathrm{ker}\,\varphi$.

Now let a BMS transformation be
\begin{align}
g=l\circ s,
\end{align}
for $l$ a Lorentz transformation and $s$ a supertranslation.
The Lorentz transformation component of a BMS transformation
$l=\varphi(g)$ is clearly well-defined.
Alternatively we can write a $g$ as
\begin{align}
g=\hat{s}\circ l
\end{align}
for $\hat{s}=l\circ s\circ l^{-1}$ some other pure supertranslation.

We can now consider composing $g_{1},g_{2}\in\mathfrak{B}$,
\begin{align}
g=g_{2}\circ g_{1}=l_{2}\circ s_{2}\circ l_{1}\circ s_{1}.
\end{align}
Because $s_{2}$ is in the normal subgroup, we can write it as conjugate to another pure supertranslation, namely
\begin{align}
s_{2}=l_{1}\circ s'\circ l_{1}^{-1}
\end{align}
for some $s'\in\mathbb{T}$. Thus
\begin{align}
g=(l_{2}\circ l_{1})\circ(s'\circ s_{1})
\end{align}
with $s'=l_{1}^{-1}\circ s_{2}\circ l_{1}$. Note that here the composition of the Lorentz transformations can be carried out via the $\mathrm{SL}(2,\mathbb{C})$ representation, whereas the composition of supertranslations is simply additive.

\bibliography{refs}

\end{document}